\documentclass[prd,twocolumn,showpacs,preprintnumbers,amsmath,amssymb]{revtex4-1}
\usepackage{graphicx}
\usepackage[usenames]{color}
\begin{document}
\preprint{BNL-96917-2012-JA}

\author{A. Bazavov}
%\email[]{obazavov@quark.phy.bnl.gov}
\affiliation{Department of Physics\\ Brookhaven National Laboratory\\ Upton, NY 11973, USA}

\author{B. A. Berg}
\affiliation{Department of Physics\\ Florida State University\\ Tallahassee, FL 32306, USA}

\title{Density of states and Fisher's zeros in compact $U(1)$ pure gauge theory }
\author{Daping Du}
\affiliation{Physics Department\\ University of Illinois\\ Urbana, IL 61801, USA}
%\email[]{daping-du@uiowa.edu}
\author{Y. Meurice}
%\email[]{yannick-meurice@uiowa.edu}
\affiliation{Department of Physics and Astronomy\\ The University of Iowa\\
Iowa City, Iowa 52242, USA }

\date{\today}
\begin{abstract}
We present high-accuracy calculations of the density of states using multicanonical methods for lattice gauge theory with a compact gauge group $U(1)$ on $4^4$, $6^4$  and $8^4$ lattices.
We show that the results are consistent with weak and strong coupling expansions. 
We present methods based on Chebyshev interpolations and Cauchy theorem to find the (Fisher's) zeros of the partition function in the complex $\beta=1/g^2$ plane.
The results are consistent with reweighting methods whenever the latter are accurate. We discuss the volume dependence of 
the imaginary part of the Fisher's zeros, the width and depth of the plaquette distribution at the value of $\beta$ where the two peaks have equal height. We discuss strategies to discriminate between first and second order transitions and explore them with data at larger volume but lower statistics. Higher statistics and even larger lattices are necessary to draw strong conclusions regarding the order of the transition. 
\end{abstract}
\maketitle

\section{Introduction}

The ongoing effort at the Large Hadron Collider has
trigerred a renewed interest in the phase structure of lattice gauge
theory models that may possibly provide an alternative to the Higgs
mechanism of the electro-weak symmetry breaking. In particular,
the location of the conformal windows
for various families of models have stirred intense discussions.
Different numerical and analytical techniques have been applied
to QCD-like models with large number of fermion flavors
\cite{appelquist09, hasenfratz09,fodor09, Deuzeman:2009mh,Fodor:2011tu,Hasenfratz:2010fi}
or with fermions in higher representations
\cite{shamir08,2010PhRvD..81k4507K,DeGrand:2011qd,Sinclair:2010be,Kogut:2011ty,DelDebbio:2010ze}. 
An interesting attempt to classify possible phases of such models
based on an effective potential for the Polyakov loop was made in \cite{Myers:2009df}.
See also \cite{DeGrand:2010ba,Ogilvie:2010vx,Sannino:2009za} for recent reviews of results and expectations.
Another direction where massive vector bosons emerge without introducing
new fermion species but in a model with modified gauge transformations
has been pursued in \cite{Berg:2010hj,Berg:2011sz}. 

In this context, it is important to understand the critical behavior of lattice models from as many consistent points of view as possible. 
Recently,  it was proposed to consider complex extensions \cite{Denbleyker:2010sv,PhysRevD.83.056009,Liu:2011zzh} of the framework  proposed by Tomboulis \cite{Tomboulis:2009zz} to explain confinement from the point of view of the Renormalization Group (RG) approach. 
 A general feature that we observed is that 
the Fisher's zeros, the zeros of the partition function in the complex $\beta$ plane \cite{fisher65}, play an important role in the determination of the global properties of the 
complex RG flows. In the case where a phase transition is present, the scaling properties of the zeros \cite{alves90b,Jersak:1996mn,Jersak:1996mj,janke00,janke01,janke04} allow us 
to distinguish between a first and second order phase transition. 

Despite its apparent simplicity, the case of the 4D pure gauge compact $U(1)$ model with  a Wilson action is not completely free of controversy. 
The presence of a double peak for the plaquette distributions near $\beta \simeq 1$ suggests a first order phase transition. 
However, if spherical or toroidal lattices are considered, the double peak disappears \cite{Jersak:1996mn,Jersak:1996mj}. In addition, 
finite size scaling, at relatively small volumes seems consistent with a second order phase transition 
with an exponent $\nu \simeq 0.35-0.40$. On the other hand, a possible scenario \cite{Klaus:1997be} is that as the volume increases, $\nu$ slowly ``rolls" towards the 
first-order value $\nu =1/D =0.25$. 
In the more recent literature \cite{Campos:1998jp,Arnold:2000hf,Arnold:2002jk,Vettorazzo:2003fg}, the idea that 
the transition is first order is favored. Using finite size scaling, the authors of Ref. \cite{Arnold:2002jk} estimated the  critical value $\beta_{\infty}=1.0111310(62)$.

In this article, we introduce new methods to locate the Fisher's zeros of the the 4D pure gauge compact $U(1)$ model with  a Wilson action. 
We rely on high accuracy determinations of the density of states, a quantity defined in Sec. \ref{sec:dos}, by multicanonical methods \cite{BergNeuhaus1991,
WangLandau2001,Bazavov:2005zy,Bazavov:2009pj} presented in Sec. \ref{sec:muca}. The lattice sizes considered are $4^4$, $6^4$ and $8^4$. 
The consistency of the results with weak and strong coupling expansion is 
checked in Sec. \ref{sec:exp}. The density of states has a convex region which implies a double peak plaquette distribution near $\beta \simeq 1$. The volume dependence of the double peak distribution is discussed empirically in Sec. \ref{sec:crossover}. 
In Sec. \ref{sec:zeros}, Chebyshev interpolations of the logarithm of the density of states and Cauchy theorem are used to find the Fisher's zeros in the complex $\beta=1/g^2$ plane. For the lowest zeros, it is possible to check consistency with reweighting methods within error bars estimated 
from the statistical fluctuations of 20 independent multicanonical streams. 

In the following, we use very high statistics on rather small lattices, because this allow us to explore new analysis methods and to test whether they converge faster towards the infinite volume limit. We plan to use similar methods for $SU(2)$ where the imaginary parts of the lowest zeros are larger and reweighting methods become less reliable when the volume increases \cite{Denbleyker:2007dy}.

Using high statistics at small volumes ($4^4$, $6^4$ and $8^4$), we show that
the imaginary part of the lowest zero appears to decrease like $L^{-3.08}$,  when the linear size $L$ increases from 4 to 8. 
This could be interpreted as signaling a second-order 
phase transition with $\nu=0.325$, a value close to the estimates of Refs. \cite{Jersak:1996mn,Jersak:1996mj}. However, using data at larger 
volumes but with lower statistics, we found indications for the ``rolling'' scenario of \cite{Klaus:1997be}. This is 
discussed in Sec. \ref{sec:largeV} where we also consider volume effects on the width and depth of the plaquette distribution at the value of $\beta$ where the two peaks have equal height. Simulations required to provide a clear cut distinction between first and second order transitions are discussed in the Conclusions. 

\section{Density of states in abelian gauge theory}
\label{sec:dos}
In the following, we consider the pure gauge partition function
\begin{equation}
Z=\prod_{l}\int \frac{d\theta_l}{2\pi} {\rm e}^{-\beta S} \  ,
\end{equation}
with $\beta\equiv1/g^2$ and the Wilson action
\begin{equation}
S=\sum_{p}(1-{\rm cos}\theta_p)\  .
\end{equation} 

We use $D$ dimensional symmetric cubic lattices with $L^D$ sites and  periodic boundary conditions. The 
number of plaquettes is 
\def\np{\mathcal{N}_p}
denoted
$\np\equiv\ L^D D(D-1)/2$.
We define the average action: 
\begin{equation}
\label{eq:pdef}
P\equiv\left\langle\mathcal{ S}/\np\right\rangle =-d(\ln Z/\np)/d\beta \ .
\end{equation}

Inserting 1 as the integral of the 
delta function over $S$  in $Z$, we can write
\begin{equation}
Z =\int_0^{2\np}dS\ n(S)\ {\rm e}^{-\beta S}\ ,
\label{eq:intds}
 \end{equation}
 with the density of states defined as 
 \begin{equation}
n(S)=\prod_{l}\int dU_l \delta(S-\sum_{p}(1-{\rm cos}\theta_p))\  .
\end{equation}
 Furthermore, we introduce the notation $s$ for $S/\np$ and we define the  entropy density $f(s)$ via the equation
 \begin{equation}
 n(S)={\rm e}^{\np  f(S/\np)} \   .
 \end{equation} 
 A more general discussion 
 for spin models \cite{alves89} or gauge models \cite{alves91} can be found in the literature where the density of states is sometimes called the spectral density.  
 From its definition, it is clear that $n(S)$ is positive. 
 Assuming that the measure for the links 
 is normalized to 1, the partition function at $\beta=0$ is 1. 
 It can be shown \cite{gluodyn04} that, if the lattice has even number of sites in each direction, and if the gauge group contains $-\openone$,  then 
 $\beta {\rm cos}\theta_p$ goes into $-\beta {\rm cos}\theta_p$ by a  change of variables $\theta_l\rightarrow \theta_l+\pi$ 
on a set of links such that for any  plaquette,  exactly one link of the set belongs to that plaquette.  Using 
\begin{equation}
	Z(-\beta)={\rm e}^{2\beta\mathcal{N}_p}Z(\beta) \ ,
	\label{eq:su2id}
\end{equation}
we find the duality  
\begin{equation}
n(2\np -S)=n(S)\  . 
\label{eq:dual}
\end{equation}
This implies the reflection symmetry 
\begin{equation}
\label{eq:mirror}
f(s)=f(2-s)\ .
\end{equation}

Numerical values of $f(s)$ have been obtained for discrete values of $s$ between 0 and 1. When $s$ is close to 0 or 2, $f(s)$ diverges logarithmically and we can only reach values of $s$ that are not too close to 0 or 2. Consequently, the results cannot be used if $|\beta|$ is too large. Using the symmetry Eq. (\ref{eq:mirror}) and interpolation methods, a continuous function can be obtained in an  interval $[\delta,2-\delta], $ where 
$\delta$ is an appropriately small quantity.  

\section{Calculation of the density of states}
\label{sec:muca}

We performed Monte Carlo simulations in pure $U(1)$ gauge theory using 
Biased Metropolis-Heatbath 
Updates \cite{Bazavov:2005zy}. To cover 
a large range of couplings $\beta\in [0,9]$ we used the multicanonical (MUCA) 
algorithm~\cite{BergNeuhaus1991} with Wang-Landau 
recursion~\cite{WangLandau2001} for
the multicanonical weights. The software we used is described in
Ref.~\cite{Bazavov:2009pj}.

We generated large statistics on symmetric lattices with volumes $4^4$, $6^4$ and
$8^4$. After the initial recursion we performed three MUCA runs on $4^4$, and
two runs on $6^4$ and $8^4$. The first MUCA run on $4^4$ was regarded as
exploratory and we did not include it in the final analysis. The weights for each
next run were refined from the previous run. In total we used 20 independent
streams for each lattice volume. In each stream we ran Wang-Landau
recursion for the multicanonical weights before the production, therefore
the weights differ between the streams, $w_{ij}(S)$, where $S$ is the total
action, $i=1,...,20$ denotes different streams and $j=1,2$ denotes
MUCA runs.

The quality of a MUCA run is indicated by the number of tunneling events
(\textit{i.e.}, how often during a run the system travels from the lowest
energy to the highest and back). Also, to estimate how many statistically
independent events we generated, we measured the integrated autocorrelation
times. These data are summarized in Appendix~\ref{app_muca}.
Our statistics consists of $N_{equi}$ sweeps for equilibration and
$N_{rpt}=64\times N_{equi}$ sweeps for measurements, where $N_{equi}=10^6$ for
$4^4$ and $6^4$ lattices and $8\times 10^5$ for $8^4$.

\begin{figure}
\includegraphics[width=3.in]{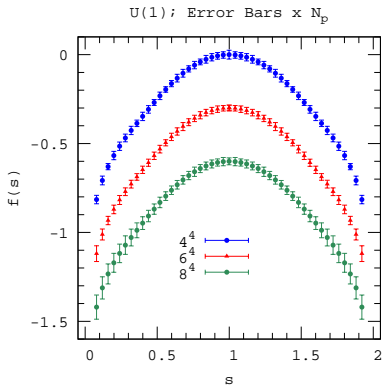}
\caption{\label{fig:fofs}$f(s)$ on a $4^4$ (top) , $6^4$ (middle) and $8^4$ (bottom) lattices; the errors have been multiplied by $\np$; for readability, arbitrary constants have been added to separate the data sets and only one of every 40 points are displayed. }
\end{figure}

For the error analysis we considered two MUCA runs in each stream
as independent measurements.
Thus, on each lattice we had 40 independent measurements in total.
For all quantities in the following the error bars are estimated from
an uncorrelated average of these 40 measurements, weighted
with the number of tunneling events in each corresponding run,
since runs with more tunnelings sample the density of states better. 
The average results for $f(s)$ are shown in Fig. \ref{fig:fofs}. 

To reweight an observable to the canonical ensemble we
need to cancel the multicanonical weight $w_{ij}(S)$ and
replace it with the Boltzmann factor $\exp(-\beta S)$.
For an observable ${\cal O}$ of interest, for instance, plaquette,
we reweight the time series accumulated during a MUCA run $ij$ to
a given coupling $\beta$:
\begin{equation}
\displaystyle
\langle O\rangle_{ij}(\beta)=
\frac{\sum_{k=1}^{N_{rpt}} O_{ij}^k\exp(-\beta S_{ij}^k) / w_{ij}(S_{ij}^k)}
{\sum_{k=1}^{N_{rpt}} \exp(-\beta S_{ij}^k) / w_{ij}(S_{ij}^k)}.
\end{equation}
The final average is then given as
\begin{equation}
\label{eq:11}
\displaystyle
\langle O\rangle(\beta)=
\frac{\sum_{i=1}^{20}\sum_{j=1}^2 N^{tunn}_{ij} \langle O\rangle_{ij}(\beta)}
{\sum_{i=1}^{20}\sum_{j=1}^2 N^{tunn}_{ij} },
\end{equation}
where the number of tunnelings $N^{tunn}_{ij}$ is given in Appendix~\ref{app_muca}.

\section{Consistency with existing results and expansions}
\label{sec:exp}
%In this section, we check the consistency of the numerical results for $f(s)$ with other results and expansions.
\subsection{Comparison with the average plaquette}
\begin{figure}
\includegraphics[width= 3.3in]{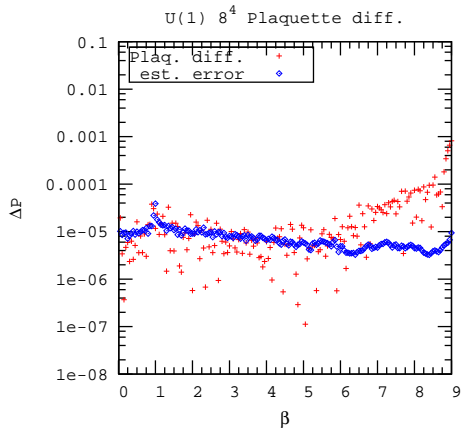}
\caption{\label{fig:plaqdiff} Difference between the average plaquette calculated with the density of states and directly,  for a $8^4$ lattice.}
\end{figure}

As a check of consistency, we compared the 
average plaquette at various $\beta$, as obtained directly from the runs, Eq. (\ref{eq:11}),  and calculated using the average density of states. 
As shown in Fig. \ref{fig:plaqdiff}, there is a good agreement within the estimated errors. 
\subsection{Series for $f(s)$}
We compared the numerical results for $f(s)$ with analytical results obtained using the weak and strong coupling expansions. The general methodology has been discussed for $SU(2)$ in \cite{Denbleyker:2008ss} and remains applicable here. The basic ingredient is the saddle point equation at $s_0$:
\begin{equation}
\label{eq:saddle}
f'(s_0)=\beta \  ,
\end{equation}
which can be used to convert an expansion of $f$ in powers of $s$ (or $(s-1)^2$) into an expansion of $s_0$ in powers of $1/\beta$ (or $\beta$, respectively). 
The coefficients of $f$ can then be determined whenever the appropriate expansion of the average plaquette is available. 
In order to take the finite volume effects into account, we need to include at least the lowest order volume correction, namely
\begin{equation}
\label{eq:finitewidth}
P=s_0+(1/2\np)(f'''(s_0)/(f''(s_0))^2) + \mathcal{O}(1/\np^2) \ .
\end{equation}
Using Eqs. (\ref{eq:saddle}) and (\ref{eq:finitewidth}) together with an existing expansion of $P$ including $1/\np$ effects up to a certain order, one can determine the coefficients of $f$ up to the corresponding order. 

It should also be noted that at finite volume, there is a finite range of $\beta$ for which $f'$ has a ``Maxwellian kink"  (discussed in Sec. \ref{sec:largeV}) and three solutions of  Eq. (\ref{eq:saddle}) are available rather than one. In this region,  both expansions are expected to fail. We now proceed to discuss them separately.

\subsection{Strong Coupling}
Following Ref. \cite{
Denbleyker:2008ss}, we define 
\begin{equation}g(y) \equiv f(1+y)\equiv  \sum_{m=0}g_{2m}y^{2m}\  .
\end{equation}
Using saddle point approximation and comparing, order by order,  with the expansion of the average plaquette from the strong coupling expansion \cite{balian74err}, we can determine the expansion of $g(y)$.  As in the case of $SU(2)$, there are logarithmic singularities at $y=\pm 1$, which can be subtracted  by defining 
\begin{equation}
h(y) \equiv g(y) -A \ln(1-y^2) \equiv \sum_{m=0} h_{2m}y^{2m} \  .
\end{equation}
The value of $A$ comes from the weak coupling expansion and will be discussed in the next subsection (see Eq. (\ref{eq:A})). 
The infinite volume results are summarized in Table \ref{tab:strong}.  The entries make clear that as the order increases, the effect of the logarithmic subtractions becomes smaller. 
This indicates singularities closer to $y=0$ ($s=1$). 
\begin{table}[h]
\newcommand\T{\rule{0pt}{2.9ex}}
\newcommand\B{\rule[-2ex]{0pt}{0pt}}
 \centering
\begin{tabular}{||c|c|c|c||}
\hline
 $m$&$a_{2m}$&$g_{2m}$&$h_{2m}$\\
 \hline
  1  &$-\frac{1}{2}$& $-1$&$-\frac{3}{4}$\\
 2  &$\frac{1}{16}$& $-\frac{1}{4}$& $-\frac{1}{8} $\\
 3 &$-\frac{13}{96}$&$\frac{43}{36}$&$\frac{23}{18}$\\
  4&$\frac{779}{6144}$&$-\frac{19}{192}$&$-\frac{7}{192}$\\
  5&$-\frac{11819}{61440}$&$-\frac{7343}{1800}$&$-\frac{7253}{1800}$\\
    6 &$\frac{2017373}{847360}$&$\frac{465331}{25920}$&$\frac{466411}{25920}$\\
	 7&$-\frac{20224291}{123863040}$&$-\frac{983357143}{1693440}$&$-\frac{983296663}{1693440}$\\
	  8\B&$\frac{5775175013}{12683575296}$&$-\frac{201757201579}{46448640}$&$-\frac{201755750059}{46448640}$\\

  \hline
\end{tabular}
\caption{\label{tab:strong}U(1) strong coupling expansion coefficients $a_{2m}$ of  $P$ (rescaled from Ref. \cite{balian74err}), and of $f(s)$ defined in the text.}
\end{table}

\begin{figure}
\includegraphics[width= 3.3in]{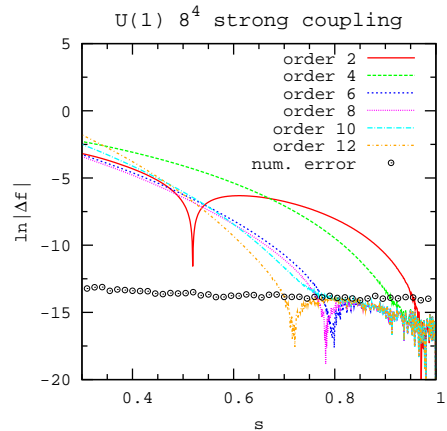}
\caption{\label{fig:errorstrong}Natural logarithm of the absolute value of the difference between $f(s)$ calculated on a $8^4$ lattice and successive approximations obtained from the strong coupling expansion. }
\end{figure}

\newcommand{\ltsim}{\protect\raisebox{-0.5ex}{$\:\stackrel{\textstyle <}
	{\sim}\:$}}
\newcommand{\gtsim}{\protect\raisebox{-0.5ex}{$\:\stackrel{\textstyle >}
	{\sim}\:$}}
The improvement of the approximation with successive order is shown in Fig. \ref{fig:errorstrong}. The graph shows that for $s\gtsim 0.5$, successive orders provide better approximations up to the 
point where the numerical accuracy is reached. 
Such a range corresponds to a convergent region $\beta\ltsim 0.9$ in the $\beta$-plane.
%This indicates that the strong coupling expansion converges for $\beta \ltsim 0.9$.

\begin{figure}
\includegraphics[width= 3.3in]{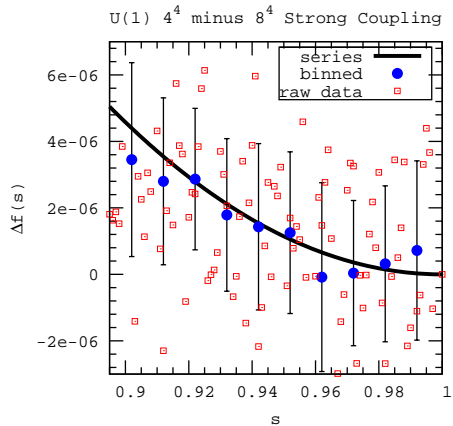}
\caption{\label{fig:errorstrongD} Difference between $f(s)$ on $4^4$ and  $8^4$ near $s$ =1 (boxes). The circles are obtained by averaging over bins of size 10. The solid line is  Eq. (\ref{eq:diffstrong}). The part with $s>1$ can be obtained by symmetry and is not shown.}
\end{figure}

It should be noted that for the strong coupling expansion, the finite volume effects are negligible for $V=8^4$. Indeed, they are even hard to resolve for $V=4^4$. This can be traced to the fact \cite{Denbleyker:2008ss} that even for this volume, the dependence on $V$ would appear at order $\beta ^8$ from the contributions 
of strong coupling graphs called torelons \cite{michael88} that wrap around the periodic volume in one direction.  
As translations in that direction do not generate new graphs, such graphs have a suppression of order $1/L$ compared to graphs with a trivial topology. 
Consequently, for order less than 8, the finite volume effects can be estimated by canceling the volume dependence in the two terms of the r.h.s. of Eq. (\ref{eq:finitewidth}). For instance at lowest order, we find that 
\begin{equation}
g_2=-1+1/(8V)\  .
\end{equation}
Consequently the difference $\Delta f(s)$ of $f(s)$ for two different volume $V_1$ and $V_2$  near $y=0$ ($s=1$) is 
\begin{equation}
\label{eq:diffstrong}
\Delta f(s) \simeq (1/8)(1/V_1-1/V_2)(s-1)^2 \ .
\end{equation}
Even for $V=4^4$, this difference is smaller than the error bars in the region $s\simeq 1$. In order to reduce the noise, we have averaged the data in bins of ten data points. 
The results are displayed in Fig. \ref{fig:errorstrongD} which shows that the data and the analytical result in Eq. (\ref{eq:diffstrong}) are compatible.

\subsection{Weak coupling}

A similar approach can be followed in the weak coupling limit. At small $s$, 
the logarithmic singularity dominates and we assume that
\begin{equation}
f(s) =  A \ln(s) + \sum_{m=0} f_m s^m \  .
\end{equation}
The unknown coefficients can be determined from the weak coupling expansion of the average plaquette 
\begin{equation}
P \simeq \sum_{m=1}b_m \beta^{-m}\  .
\end{equation}
The volume dependent coefficients $b_m$ have been calculated up to order 4 in Ref. \cite{Horsley:1981gj}. 
The two lowest orders of the expansion can be performed exactly and yield: 
\begin{eqnarray}
\label{eq:A}
A&=&1/4-5/(12V)\ ,  \\
f_1&=&(1/8)(1-1/V) \ .
\end{eqnarray}
The higher orders involve numerical loop calculations. The results of the expansion as well as the volume corrections for the $4^4$ and $6^4$ lattices are shown in Table \ref{tab:weak}.
\begin{table}[h]
\newcommand\T{\rule{0pt}{2.9ex}}
\newcommand\B{\rule[-2ex]{0pt}{0pt}}
\centering
\begin{tabular}{||c|c|c|c||}
\hline
  \T &$V=\infty$&$V=6^4$&$V=4^4$\cr
 \hline
$b_1$ \T&  $\frac{1}{4}$&  $\frac{1295}{5184}$ &  $\frac{255}{1024}$ \cr
$b_2$ \T&  $\frac{1}{32}$&  $\frac{2171747375}{208971104256}$&  $\frac{65025}{2097152}$\cr
$b_3$ \T&  0.01311& $0.01309$& $0.01296$\cr
$b_4$ \T&  0.00752& $0.00749$& $0.00739$\cr
\hline

 \hline
$A$ \T&  $\frac{1}{4}$&$\frac{3883}{15552}$ &  $\frac{763}{3072}$\cr
$f_1$ \T& $\frac{1}{8}$&$\frac{1295}{10368}$&  $\frac{255}{2048}$\cr
$f_2$ \T&  0.07363& $0.07359$ &  $0.07314$\cr
$f_3$ \T&  0.07638& $0.07605$ &  $0.07515$\cr
\hline
\end{tabular}
\caption{\label{tab:weak}The weak coupling expansion for $V=\infty$,  $4^4$ and $6^4$. The upper half is the list of the expansion coefficients of the average plaquette $P$ with respect to $1/\beta$ \cite{Horsley:1981gj}. The lower half is the corresponding list of expansion coefficients of $f(x)$. } 
\end{table}

\begin{figure}
\includegraphics[width= 3.3in]{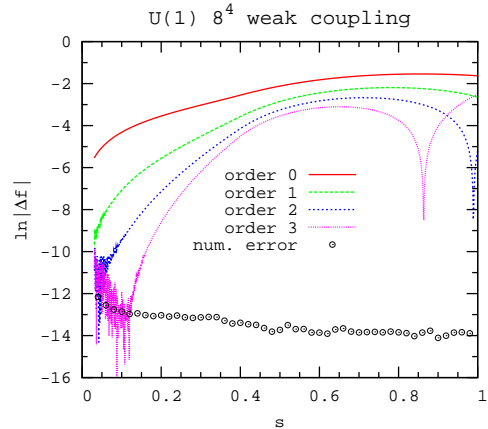}
\caption{\label{fig:errorweak}   Difference between $f(s)$ calculated on a $8^4$ lattice and successive approximations obtained from the weak coupling expansion.}
\end{figure}

The difference between numerical and analytical results is shown in Fig. \ref{fig:errorweak} for $V=8^4$. 
The graph makes it clear that the quality of the approximation increases as three successive corrections to the 
leading logarithm are added. The third correction is good enough to reproduce the data within the numerical accuracy for $s<0.1$. 
This order is not sufficient to identify a ``non-perturbative envelope" defined in Ref.  \cite{npp} and observed for $SU(2)$ in Ref.  \cite{Denbleyker:2008ss}.

\begin{figure}
\includegraphics[width= 3.3in]{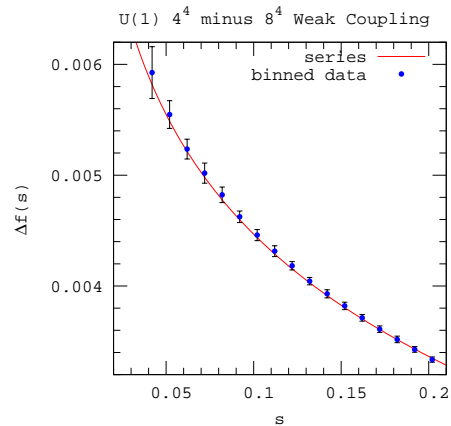}
\caption{\label{fig:errorWD}  Difference between $f(s)$ on $4^4$ and  $8^4$ near $s$ = 0 with average over bins of size 10. The solid line is the expansion described in the text.}
\end{figure}

The difference between $V=4^4$ and $V=8^4$  is shown in Fig.  \ref{fig:errorWD}. 
Note that for the smallest volume ($V=4^4$), the resolution in $s$ used during the multicanonical simulation 
is coarser than the 1,000 bins used to represent $f(s)$. Consequently, some small ``staircase'' structure appears near 0 where $f(s)$ changes rapidly. 
For this reason, we have averaged $f(s)$ over bins of size 10 and Fig.~\ref{fig:errorWD} shows a good agreement with the analytical expansion that includes the logarithmic singularity 
and a linear term. Higher order corrections are significantly smaller than the errors bars. 
There is an arbitrary constant in the expansion of $f(x)$ which cannot be determined by the saddle point equation. For the numerical data, such a constant may differ for two different volumes and needs to be subtracted.

\section{Volume dependence in the crossover region}
\label{sec:crossover}

\subsection{Empirical parametrization}

\begin{figure}
\includegraphics[width= 3.3in]{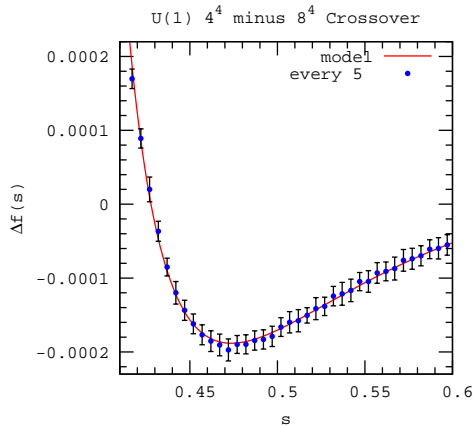}
\caption{\label{fig:difffM} Difference between $f(s)$ on $4^4$ and  $8^4$ near $s$ = 0.5. For readibilty, we only show every 5 points (no binning). The solid line is the fit given in Eq. (\ref{fig:hydro}).}
\end{figure}

The difference of $f(s)$ for 
$4^4$ and $8^4$ resembles the effective potential for the central Coulomb potential with a leading singularity near $s$ = 0.35 and a $1/s$ behavior at larger $s$. 
Using in addition a constant that has no particular meaning as long as we don't normalize the density of states and a $1/s^2$ correction, we performed a 4-parameter fit 
with the 311 bins corresponding to $0.39<s<0.7$. The numerical result is:
\begin{eqnarray}
\label{fig:hydro}
f(s)=&-&0.00112063+4.82641\times 10^{-6}/(-0.35+s)^2 \cr &-&0.000680501/s^2+0.00172882/s .
\end{eqnarray}
As shown in Fig. \ref{fig:difffM}, it fits the data reasonably well. 

\subsection{Volume dependence of the double peak}

\begin{figure}
\includegraphics[width=3.3in]{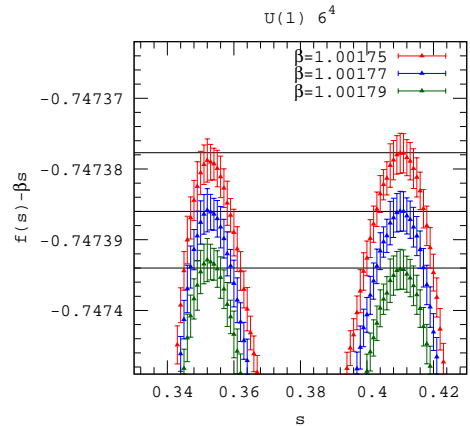}
\caption{\label{fig:dpeak6to4} $f(s)-\beta s$ for  $\beta$ = 1.00175, 100177 and 1.00179 on a $6^4$  lattice.  The horizontal lines have been drawn to emphasize small height asymmetries.}
\end{figure}

The plaquette distributions for the volumes  considered here have a double peak structure for $\beta$ near 1. 
%If the double-peak persists at arbitrarily large volume, we are in presence of a first-order phase transition. 
At finite volume, it is easy to locate the value of $\beta$, denoted $\beta_S$ hereafter,  where the two peaks of $f(s) -\beta_S s$ have equal height.  
%At finite volume, this quantity might not be a completely accurate estimate of the value of $\beta$ for which the average action changes abruptly because the equality of height does not guarantee the equality of widths, however  it is easy to determine this quantity numerically.   
Other pseudocritical $\beta$ have been defined in the literature \cite{1991JSP....62..529B,1990JSP....61...79B,Billoire:1992ke,Klaus:1997be}.
The accuracy of the determination of $\beta_S$ depends on the smoothness of the distribution and the size of the error bars. In Fig. \ref{fig:dpeak6to4}, we show that $f(s)-\beta s$ is slightly tilted to the left for $\beta$ = 1.00175 and to the right for $\beta$ = 1.00179. Given the smoothness of the distribution, we conclude that $\beta_S=1.00177(2)$. 
With the same graphs, we can also determine approximate values of the two maxima $s_1$ and $s_2$.  The numerical results for the three volume considered are provided in Table \ref{tab:betas}. 
\begin{table}
\begin{center}
\begin{tabular}{||c|c|c|c||}
\hline
$L$ & $\beta_S$ &$s_1$ & $s_2$\cr
\hline
4&0.9793(1)&0.370(5) &0.445(5)\cr
6&1.00177(2)&0.353(2)&0.411(2)\cr
8&1.00734(1)&0.349(1)&0.395(1)\cr
\hline
\end{tabular}
\end{center}
\caption{\label{tab:betas} $\beta_S$, $s_1$ and $s_2$ defined in the text for $L$ = 4, 6 and~8.}
\end{table}

\begin{figure}
\includegraphics[width=3.3in]{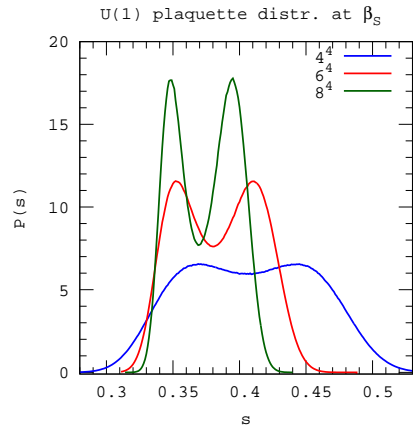}
\caption{\label{fig:dpeak}Plaquette distribution for $U(1)$  at $\beta_S$ for $L$= 4, 6 and~8.}
\end{figure}

The density of states can be used to calculate the plaquette probability distribution at  $\beta_S$.  The results are shown in Fig. \ref{fig:dpeak}  where the normalization has been chosen in such a way that the integral under the curve is approximately one. Fig. \ref{fig:dpeak} makes it clear that the dip between the peaks deepens and the peak separation decreases as the volume increases. 
%The peak separation $s_2-s_1$ scales approximately like $L^{-0.7}$. Given that we have only three data points, this statement should be used to estimate the range 
%of $\beta$ necessary to calculate the density of states at slightly larger volume rather than a statement about what happens at infinite volume.  
This will be discussed in more detail in Sec. \ref{sec:largeV}.

\section{Fisher's zeros}
\label{sec:zeros}
\subsection{Approximate zeros from reweightings}

Approximate values of $Z$ at fixed $\beta$ can be obtained by using the Riemann sum approximations of 
Eq. (\ref{eq:intds}):
\begin{equation}
\label{eq:dsum}
Z(\beta)\simeq \Delta s\sum_s{\rm e}^{\np(f(s)-\beta s)}\ .
\end{equation}
We can now study how the error $\delta f$ on $f(x)$ can affect our estimates of $Z$. 
The relevant quantity is $\np \delta f$ is included in Fig. \ref{fig:fofs}. For the three volumes, $\np \delta f$ is of the order of a few percents and linearization is justified. Small scale fluctuations of the same order are visible in the 
distributions of the independent streams. 

\begin{figure}
\includegraphics[width=4.3in]{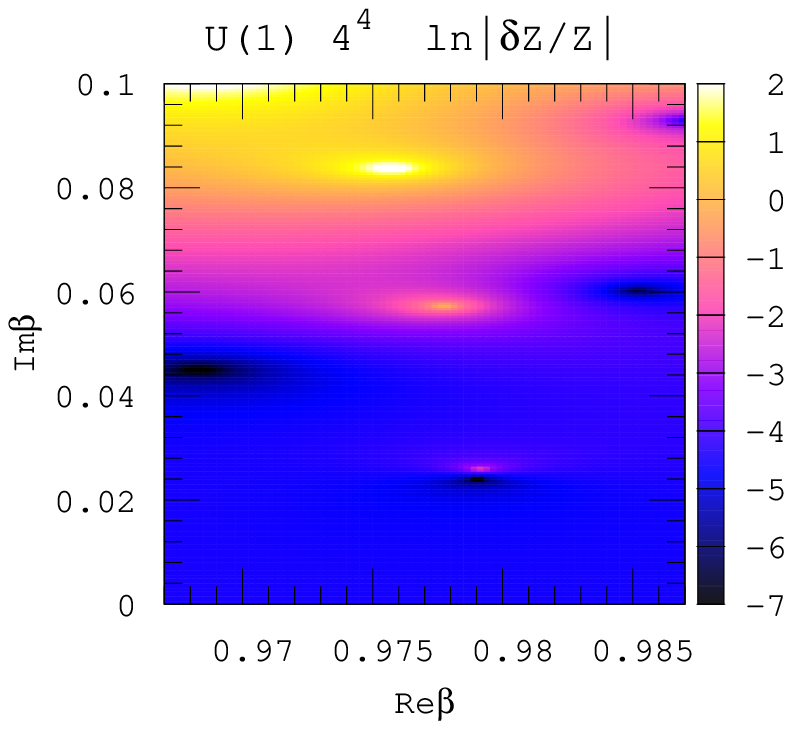}
\includegraphics[width=4.3in]{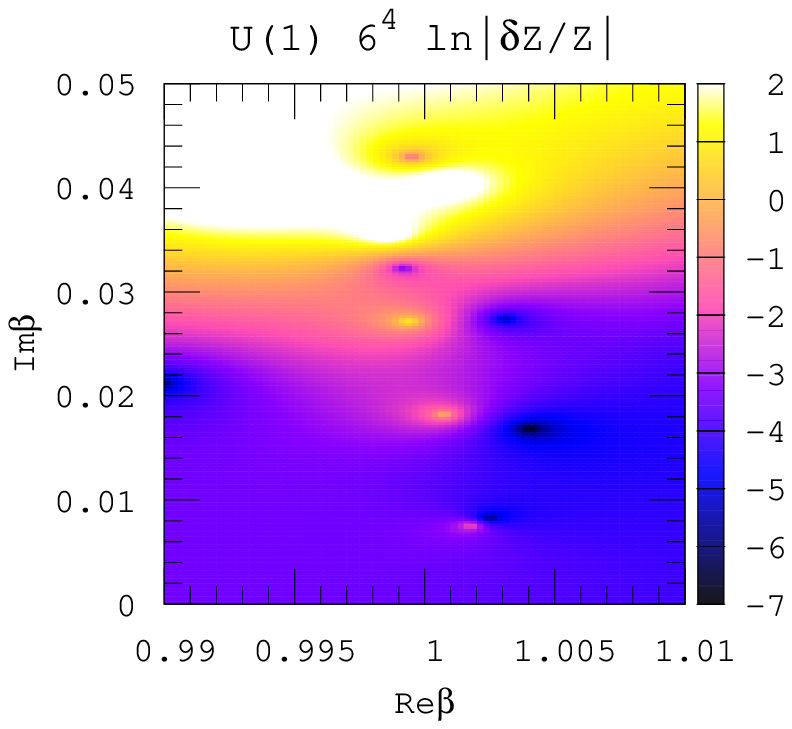}
\caption{\label{fig:contour} ln$|\delta Z/Z|$ for $4^4$ and $6^4$ lattices (color online).}
\end{figure}

As we are interested in locating Fisher's zeros, it is clear that the 
errors have a potentially important effect near an approximate zero. 
The best we can do is to identify regions where $|Z|$ is significantly  larger than $|\delta Z|$ so that we can confidently say that there are no zeros in these regions. 
If we use the linear estimate
\begin{equation}
\delta Z(\beta)\simeq \Delta s\sum_s\np \delta f(s) {\rm e}^{\np(f(s)-\beta s)}\ ,
\end{equation}
we have the inequality 
\begin{equation}
|\delta Z(\beta)|< \Delta s\sum_s\np |\delta f(s)| {\rm e}^{\np(f(s)-Re\beta s)}\ ,
\end{equation}
but in general the bound is not sharp because the sign of $\delta f(s)$ can vary rapidly. 
We have estimated $|\delta Z|$ by taking the difference 
between $Z$ calculated with the averaged $f$ and $Z$ calculated  with the stream with the most tunnelings. The results are shown in
Fig. \ref{fig:contour}. A mesh of 0.00025 in $\beta$ is used which is larger than the typical fluctuations in $f$. The light (toward yellow on-line) regions represent the areas where we cannot exclude zeros. 
The dark (toward blue on-line) regions represent the areas where zeros are very unlikely.
A small light region is an indication for the existence of a zero while a broad light region indicates that the errors dominate. The second possibility typically appears at large imaginary 
$\beta$ where due to rapid oscillations of the integrand, cancellations occur making the final results more sensitive to the errors on $f(s)$.  In view of this remark, Fig. \ref{fig:contour} suggests that reweighting methods allow to estimate the locations of the two lowest zeros for $L=4$ and three lowest zeros for $L=6$. 

\begin{figure}
\includegraphics[width=3.6in]{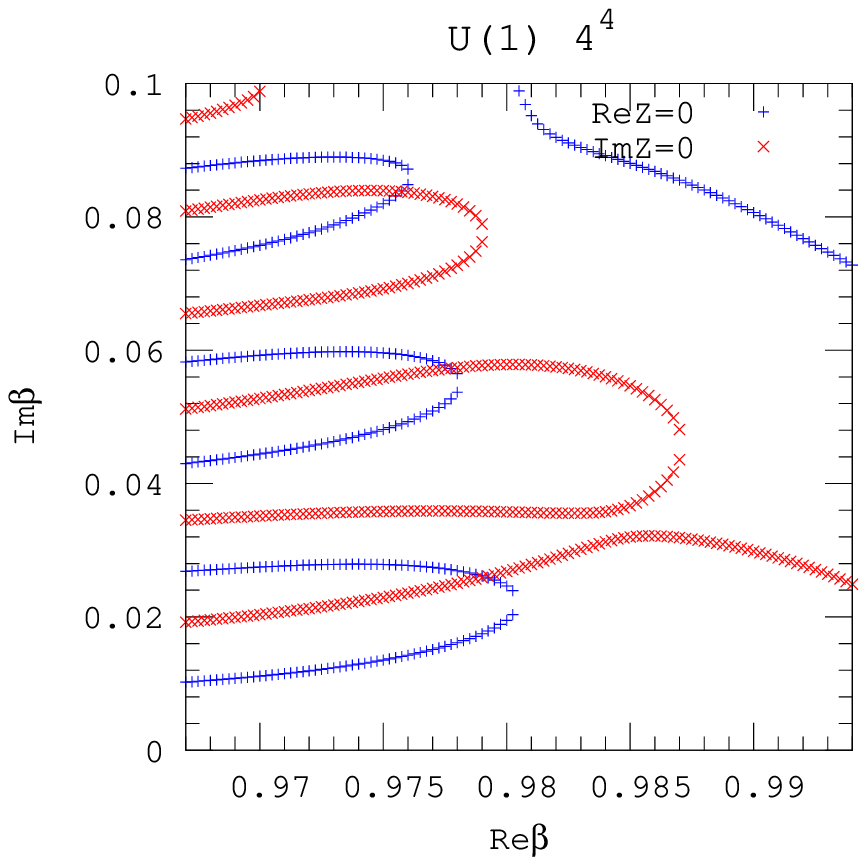}
\includegraphics[width=3.6in]{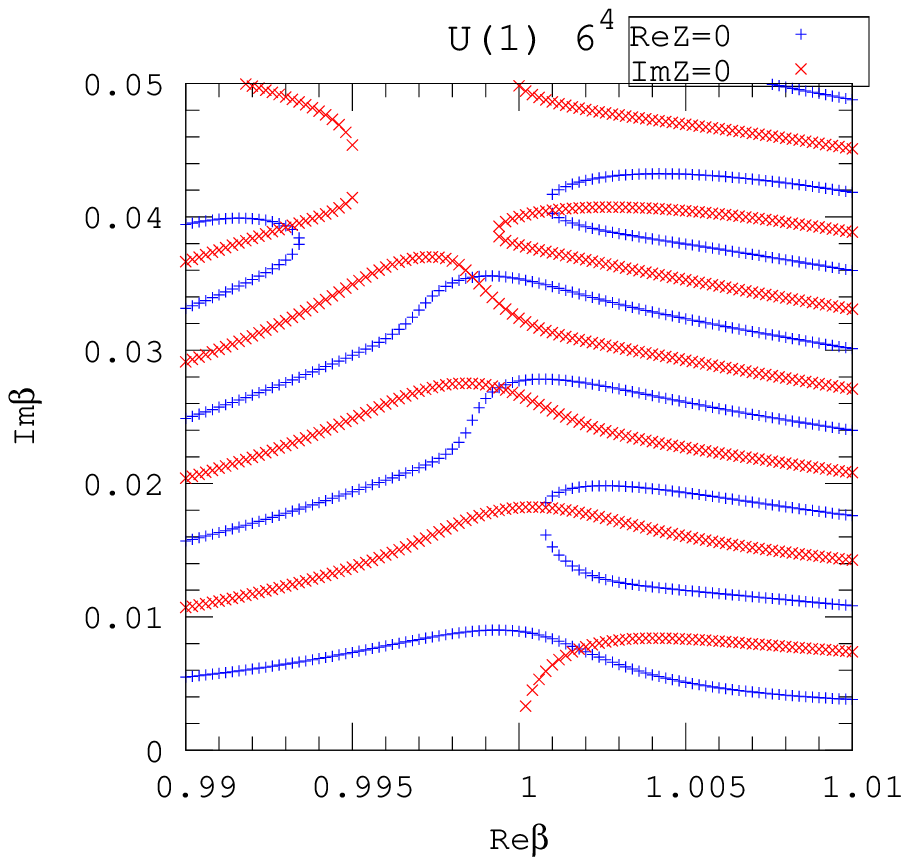}
\caption{\label{fig:rewz}Zeros of the Re (+, blue on-line) and Im (x, red on-line) part of $Z$ for $U(1)$ using the density  of states for $4^4$ and $6^4$ lattices.}
\end{figure}

We have also calculated Re$Z$ and Im$Z$ from Eq. (\ref{eq:dsum}) using the average $f$. Their respective zeros are shown in Fig. \ref{fig:rewz}. 
The complex zeros appear at the intersections of the two sets of curves defined by Re$Z$=0 and Im$Z$=0 respectively. This happens in a way which is consistent with Fig. \ref{fig:contour}.  Error bars can be estimated by comparing 
the intersections for the 20 streams. The results are given in Tables \ref{tab:re} and \ref{tab:im}. 

\begin{table}
\begin{center}
\begin{tabular}{||c|c|c|c||}
\hline
$L$ & {\rm first} &{\rm second} & {\rm third}\cr
\hline
4&0.9791(1)&0.9780(4)&not\  stable\cr
6&1.00180(5)&1.0007(1)&0.9993(5)\cr
8&1.00744(2)&1.0068(2)&1.0061(4)\cr
\hline
\end{tabular}
\end{center}
\caption{\label{tab:re} Real part of the first three zeros.}
\end{table}

\begin{table}
\begin{center}
\begin{tabular}{||c|c|c|c||}
\hline
$L$ & {\rm first} &{\rm second} & {\rm third}\cr
\hline
4&0.0259(1)&0.057(1)&not\ stable\cr
6&0.00758(2)&0.018(1)&0.027(2)\cr
8& 0.00306(2)&0.008(1)&0.012(1)\cr
\hline
\end{tabular}
\end{center}
\caption{\label{tab:im}Imaginary part of the first three zeros.}
\end{table}

\subsection{Chebyshev interpolations}

The original grids of the density of states are sometimes not sufficient for precise numerical integrations which is how we define our partition function. It is especially true when the imaginary component of $\beta$ is large and, as a consequence, the partition function oscillates more frequently than the original grids can resolve. It is convenient to apply the Chebyshev interpolation which provides arbitrary integrating step sizes for designed integral precision. For the Chebyshev interpolation of numerical data, the determination of the coefficients by the Least Square Fit method is more efficient and robust than by discrete or integration methods. In this paper, we will primarily follow this approach. 

A range of interest $[a, b]$ can be mapped to $[-1, 1]$ in which we express the target function by
\begin{equation}
\label{eq:cexp}
f(y) = \sum_{n=0}^{N_c} c_n T_n(y) \ .
\end{equation} 
We then minimize the distance of the function to a data set or multiple data sets, which will uniquely determine the coefficients $c_n$ of linear equations. 

We shall keep in mind that, like other polynomial approximations, Chebyshev interpolations may introduce artifacts such as fake zeros. We want to make sure that the true zeros are distinct from the fake ones. Special attention should be paid to the range of approximation. In practice, we often use a small range to emphasize the numerical signal from a certain region. The average plaquette $\langle x \rangle$ is related to the coupling $\beta$, through $\langle x \rangle _\beta = -\partial \ln Z(\beta) /\partial \beta /\np$. 
This is not valid if $\langle x \rangle$ goes beyond the range of the approximation (an ellipse in the complex plane, see below).
% and the corresponding $\beta$'s (zeros) are not reliable. 
Reducing the range of interpolation may introduce fake zeros with large Im$\beta$ in the $\beta$ complex plane. However the lowest zeros are usually not affected. Care should also be paid on the orders of the Chebyshev approximation. True zeros should be independent of the order of polynomials. In the following, we always use various orders of Chebyshev interpolations and make sure that the zeros are free of these artifacts. 

\subsection{Ellipse of convergence}

The definition 
\begin{equation}
T_n(z)=\cos (n \ {\rm arcos} (z))
\end{equation}
shows that the expansion in Chebyshev polynomials is a Fourier expansion for the variable $\arccos(x)$.  
If $|c_{n+1}/c_n|$ from Eq. (\ref{eq:cexp}) reaches a limit $C$, then the expansion converges for $|T_n(z)|<C^{-n}$. 
To work on the complex plane, the following relation is helpful: $T_n(z) = (\omega^n + \omega^{-n})/2$, when $z$ is expressed as $z=(\omega + \omega^{-1})/2$. The convergence of a Chebyshev series is then analyzed through the variable $\omega$. It can be shown \cite{Boyd:1988} that the region of convergence on the $\omega$-plane is a ring confined by a pair of concentric circles and the region is mapped into an area bounded by an ellipse on the $z$-plane.   

The continuation of the Chebyshev expansion to the complex plane is limited by the ellipse. Fortunately, in the case of $U(1)$, the lowest complex zeros are typically very close to the real axis and these zeros are well inside the ellipse of convergence. 

\subsection{Locating zeros with the residue theorem}

There is a general algorithm to find the zeros of an analytic function by using Cauchy's Integral Theorem \cite{124403}. For simplicity, we will only consider the special case when all the zeros are of order 1 which apparently applies to our problem. Suppose that an analytic function $Z(\beta)$ has $K$ zeros enclosed by a closed contour $C$, then 
\begin{equation}
\frac{1}{2\pi i}\oint_c (\ln Z)'\; \beta ^n\; d\beta = \sum_{i=1}^K (\beta_i)^n, \;\;n=0,1,2,... \
\label{eq:cauchy}
\end{equation} 
where $\beta_i$ are all the zeros in contour $C$. When $n=0$, the summation on the right hand side is just the number of zeros. 

The partition function we are considering is an analytic function, since it is just a sum of analytic functions. 
%Theoretically, we could construct a large contour which encloses $K$ zeros, then perform 
%$K$ integrations in Eq. (\ref{eq:cauchy}) and solve the resulting $K$-order equations. 
%However %higher order nonlinear equations are extremely sensitive to the error of the coefficients. 
%So practically, we will only work with contours with at most two zeros. 
We scan the complex plane with rectangular contours which enclose two or less zeros. We monitor the $n=0$ integral which should give the number of zeros very close to an integer and a very small imaginary part. The method turns out to be quite robust and reliable.  
 
\begin{table}
\begin{center}
\begin{tabular}{||c||c|c|c||c|c|c||}
\hline
$L$ & Re$\beta$ & $\sigma_{s}$& $\sigma_{c}$ & Im$\beta$ & $\sigma_{s}$ &$\sigma_{c}$\cr
\hline
 &0.9791235 &3.6e-5&5.3e-8&0.0260065& 3.7e-5&3.9e-9\cr
4 &0.9777314 &3.5e-4&7.1e-6 & 0.0572764&  1.4e-4&3.3e-6\cr
 & 0.9752954 &1.1e-3	&2.9e-4 &0.0831705&  1.3e-3&3.2e-4 \cr
\hline
& 1.0017969&1.7e-5&1.7e-6&0.0075821&8.7e-6&1.4e-6\cr
6 & 1.0007433 &6.0e-5&2.3e-5&0.0182044&2.8e-5	&4.0e-6 \cr
 & 0.9988964&1.4e-4&2.7e-4	&0.0271866&4.5e-4&1.5e-4\cr
\hline
 & 1.0074380&1.1e-5&7.7e-8&0.0030653&3.6e-6&6.8e-8\cr
8 & 1.0068296	&2.3e-5&2.1e-6&0.0077673&2.4e-5&3.3e-7	 \cr
 &1.0060410&1.1e-4&1.2e-5&0.0115079&1.0e-4&8.5e-5\cr
\hline
\end{tabular}
\end{center}
\caption{\label{tab:compare}  The lowest three zeros in the volumes $4^4$, $6^4$ and $8^4$. Columns 2-4 are  the real parts of the zeros, the estimate error $\sigma_s$ from different streams of Monte Carlo runs and the error $\sigma_c$ due to the orders of Chebyshev interpolation (we used three different orders $40$, $44$ and $50$ for all three volumes). Columns 5-7 are similar quantities for the imaginary parts. }
\end{table}
\subsection{Zero structure near the real axis}

\begin{figure}
\includegraphics[scale=0.8]{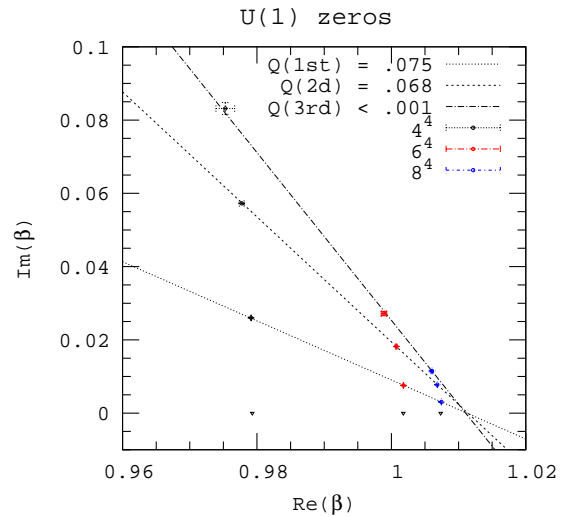}
\caption{ \label{fig:3zeros}The lowest zeros from the volumes $4^4$, $6^4$ and $8^4$ (from left to right). Linear fits for the first, second and third zeros (bottom to top) and their goodness of fit $Q$. The diamonds on the real axis are the double-peak $\beta$'s from Table \ref{tab:betas}. }
\end{figure}

The lowest zeros from three volumes are  given in Table \ref{tab:compare} and shown in Fig. \ref{fig:3zeros}. The error bars take into account  both the Monte Carlo statistical error and the (much smaller) Chebyshev interpolation error. The three lines are the linear fits for the first, second and third lowest zeros. They intersect the real axis approximately at the same point $\beta=1.01134(1)$.  Fig. \ref{fig:3zeros}  also  shows that $\beta _S$ and the real part of the zeros are highly correlated. 

The good look of the linear fits is deceptive as they have a rather large $\chi^2$ and a small goodness of fit $Q$  (see p. 111 
of \cite{BBbook}) which can be explained by the small errors bars. 
Another potentially deceptive result is that the imaginary part of the lowest zero decreases like $L^{-3.08}$. If this result is indicative of what happens 
at larger volume, this would be interpreted as signaling a second order phase transition with $\nu \simeq 1/3.08 \simeq 0.325$. Larger lattices are needed as will be discussed in Sec. \ref{sec:largeV}.

\subsection{Dependence on the range of integration}

In the previous calculations, the tails of integration play a marginal role. If we are interested only in the zeros near $\beta =1$, the density of states in a finite range is sufficient. This information becomes very important at higher volumes where the calculation is more expensive. In Table \ref{tab:range}, we 
provide the values $s_a$ and $s_b$ outside of which the knowledge of $f(s)$ has effects smaller than the error bars on the zeros.  
\begin{table}
\begin{center}
\begin{tabular}{||c|c|c|c|c||}
\hline
$L$ & $s_a$&$s_b$ &$\beta_a$ & $\beta_b$\cr
\hline
4   &  0.274   &  0.488   &   1.125  & 0.945\cr
6  &  0.284  & 0.436 & 1.1 & 0.985\cr
8 &   0.295  & 0.408 &1.075 &1\cr
\hline
\end{tabular}
\end{center}
\caption{\label{tab:range} Values of  $s_a$ and $s_b$ and the corresponding values of $\beta$ ($P(\beta)=s$). }
\end{table}

\section{Toward larger volume calculations and scaling}
\label{sec:largeV}

In this section, we explore ways of discriminating between first and second order transitions. For this purpose, we  
include data at larger volumes ($L$=10, 12, 14 and 20) from Ref. \cite{Berg:2006hh} with much lower statistics, 
namely one stream with two MUCA runs. 

\subsection{Zeros}

In Sec. \ref{sec:zeros}, we explained that for 
the data for $L=4$, 6 and 8, the imaginary part of the lowest zeros scales like $L^{-3.08}$.  
It is possible that as the volume increases, the approach of the real axis ``rolls'' toward the $L^{-4}$ scaling expected for a first order transition \cite{Klaus:1997be}. We now discuss the scaling of the zeros using 
the lower statistics data for the larger volumes given in Table \ref{tab:betH}. 
\begin{table}
\begin{center}
\begin{tabular}{||c|c|c|c|c||}
\hline
$L$ & Re$\beta$ & $\Delta Re/2$& Im$\beta$ & $\Delta Im/2$\cr
\hline
10 &1.00947 & 2e-5& 0.001478&  2e-6\cr
12 &1.01027&  2e-5 &0.000795& 2e-6\cr
14& 1.01064 & 2e-5& 0.000449& 8e-6\cr
20 &1.01101  &1e-5& 0.000119 &1e-6\cr
\hline
\end{tabular}
\end{center}
\caption{\label{tab:betH} Higher volume zeros. Columns 2 and 4 are the averages over the two MUCA runs. Columns 3 and 5 are one half of the differences (not the estimated error, see text). }
\end{table}

It is questionable that two MUCA runs could lead to a reliable estimate of the errors.  An error bar from just two independent measurements fluctuates 
strongly and reaches a 95\% confidence range only at about 14 (instead
of 2) error bars (see p.78 of \cite{BBbook}). We decided therefore to 
smoothen the error bars by assuming that the real relative error is 
the same for all four of our large lattices. Averaging these relative 
errors and multiplying the by three, the approximate 95\% confidence 
range of four independent data, gives an error bar of 1.69\%, which 
is then given in the fourth column for all the large data of of 
Table~\ref{tab_ImzL20}. Not to overweight the far more accurate
small lattice against the large lattice data in the subsequent 
fits, they are also used with a relative error of 1.69\% and
thus listed in Table \ref{tab_ImzL20}.  We want to emphasize that this procedure has been 
designed to understand how different fits allow to discriminate between first and second order 
rather than to extract accurate values for the fitted parameters. 

\begin{table}[th] \begin{center}
\begin{tabular}{|c|c|c|c|} \hline
$L$ & First Run & Second Run &  Combined \\ 
\hline
  4 & --          & --          & 0.02691    (44)  \\ \hline
  6 & --          & --          & 0.00758    (13)  \\ \hline
  8 & --          & --          & 0.003065   (52)  \\ \hline
 10 & 0.0014756   & 0.0014797   & 0.001478   (25)  \\ \hline
 12 & 0.0007927   & 0.0007969   & 0.000795   (14)  \\ \hline
 14 & 0.00045747  & 0.00044157  & 0.0004495  (76)  \\ \hline
 20 & 0.000011882 & 0.000011901 & 0.00001189 (21)  \\ \hline
\end{tabular}
\caption {\label{tab_ImzL20} $y={\rm Im}z$ from two 
independent runs on $L\ge 10$ lattices and their combination
as explained in the text together with reduced accuracy values
from $L\le 8$ lattices.} \medskip
\end{center} \end{table} 

For these data we performed 3-parameter fits, which are listed 
in the following together with their goodness of fit $Q$.
\begin{eqnarray} \label{3p4w1ox2}
  y &=& \frac{a_1}{L^4}\,\left(1+\frac{a_2}{L}+\frac{a_3}{L^2}\right)
  \,, ~~~Q=0.43\,,\\ \label{3p1ox}
  y &=& a_1L^{a_2}\,\left(1+\frac{a_3}{L}\right)\,,
  ~~~Q=6.2\times 10^{-4}\,, \\ \label{3par4}
  y &=& \frac{a_1}{L^4}\,\left(1+a_2L^{a_3}\right)\,,
  ~~~Q=2.8\times 10^{-3}\,.
\end{eqnarray} 
The first fit shows that $L^{-4}$ behavior is consistent with all
the data put together. The other two fits are in disagreement with
the data.

\begin{figure}% \begin{center}
\includegraphics[width= 3.3in]{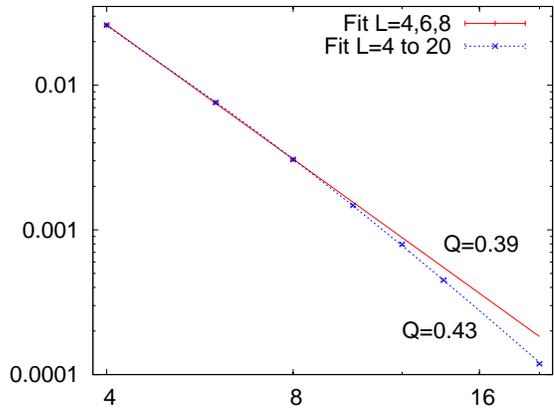}
\caption{Fits of ${\rm  Im} z (L)$ on a log-log scale.} \label{fig_U1lnln}
%\end{center}
 \end{figure}

Using {\it only} the data from the $L=4,\,6,\,8$ lattices with the {\it modified} error bars given in Table \ref{tab_ImzL20}, the 2-parameter fit
\begin{eqnarray} \label{2p}
  y &=& a_1\,L^{a_2}\,,~~~Q=0.39
\end{eqnarray} 
is also in agreement with the data and gives the exponent 
$a_2=-3.082\,(35)$ instead of $-4$. The fits (\ref{3p4w1ox2}) and (\ref{2p}) are shown in Fig. \ref{fig_U1lnln}.

However, the fit 
\begin{eqnarray} 
\label{more}
  y &=& \frac{a_1}{L^{3.08}}\,\left(1+\frac{a_2}{L}+\frac{a_3}{L^2}\right)
 \ ,
\end{eqnarray} 
with the seven data points leads to $Q<10^{-8}$. In addition, if we perform a four parameter fit as in Eqs. (\ref{3p4w1ox2}) and (\ref{more}) but with the leading exponent fitted, we obtain 4.121(74) for this exponent with $Q$=0.72. These results seem to favor the first order possibility. However, they should be checked with 
higher statistics data for the larger volumes. 

\subsection{Features of $f(s)$}

\begin{figure}
\includegraphics[width= 3.3in]{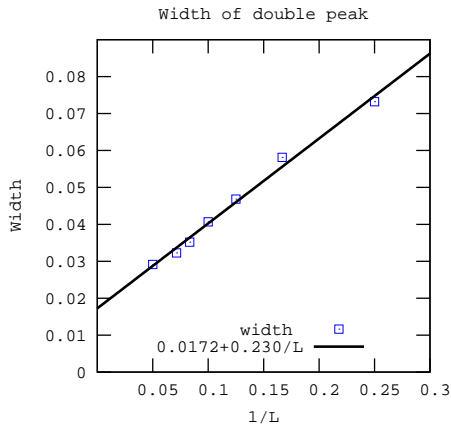}
\includegraphics[width= 3.3in]{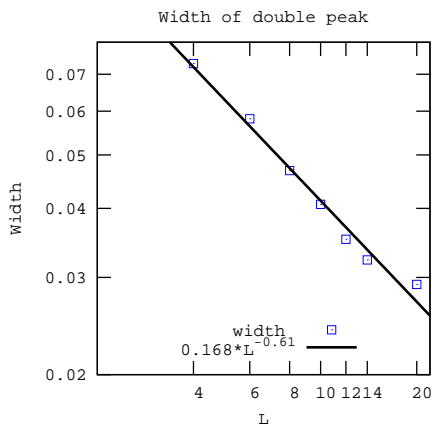}
\caption{\label{fig:width} Width of $ f-\beta_S s$ as a function of $L$. }
\end{figure}

In the infinite volume limit, the width of the double peak distribution goes to a nonzero limit (latent heat) for a first order phase transition. 
For a second order transition, this width should go to zero as an inverse power of $L$. These two possibilities are tested by plotting the 
width versus $1/L$ or in a log-log scale in Fig. \ref{fig:width}. 

\begin{figure}
\includegraphics[width= 3.3in]{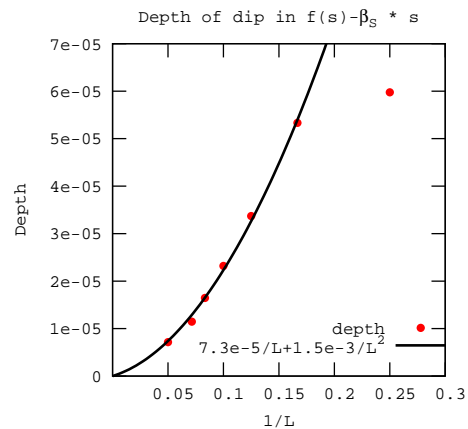}
\includegraphics[width= 3.3in]{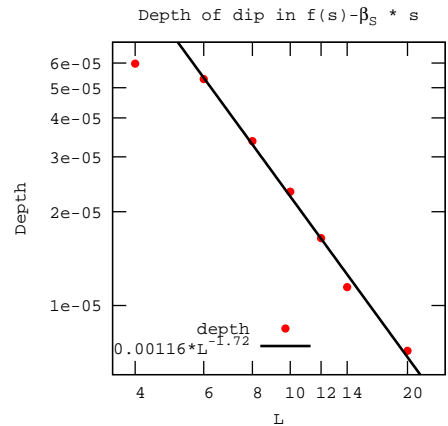}
\caption{\label{fig:depth} Difference between the local minimum and the local maxima of $ f-\beta_S s$ as a function of $L$. }
\end{figure}

The difference between the local minimum and the local maxima of $ f-\beta_S s$ (by definition of $\beta_S$, the two local maxima have the same height) 
should decay like $C/L$ for a first order transition with $C$ proportional to the interface tension. 
For a second order transition, this difference should go to zero as an inverse power of $L$.
The data is shown 
versus $1/L$ and 
on a log-log scale together with simple fits (made without $L=4$) in Fig. \ref{fig:depth}. 
In the first fit, the $1/L^2$ corrections are clearly important and it is not surprising that the power in the 
second fit is between -1 and -2. 

Using arguments from Ref. \cite{PhysRevE.74.011108}, leads to the conclusion that for a second order phase transition, the width should scale like 
$L^{-(1-\alpha)/\nu}$, while the depth should scale like $L^{-D}$. If we use $D=4$, $\nu\simeq 0.325$ and the hyperscaling relation $\alpha=2-D\nu\simeq 0.7$, 
we obtain $(1-\alpha)/\nu \simeq 0.92$ which is not too far off for the width but very far off for the depth.  

\section{Conclusion}
\label{sec:concl}
Using multicanonical methods we have calculated the density of states for pure $U(1)$ lattice gauge theory with high precision on small $4^4$, $6^4$ and $8^4$ lattices and with moderate precision on larger $10^4$, $12^4$, $14^4$ and $20^4$ lattices. From these data we were able to locate low-lying Fisher's zeros by Chebyshev interpolations and residue theorem methods. On the small lattices the scaling properties of the zeros are consistent with a second order phase transition, while from the larger lattices there is some indication that this turns around and becomes consistent with a first order transition. 

Although $U(1)$ lattice gauge theory was already introduced in the pioneering paper by Wilson \cite{wilson74c}, it still resists to reveal clearly the true nature of its transition from the confinement to the Coulomb phase. Like other physical quantities, e.g., Polyakov loop susceptibilities, Fisher's zeros appear to need rather large lattices to display their asymptotic scaling properties. As modern supercomputers allow parallel processing on an unprecedented scale, the solution may finally become achieved by brute force calculations on very large lattices.

\begin{acknowledgments}
This 
research was supported in part  by the Department of Energy
under Contracts No. FG02-91ER40664, DE-FG02-97ER41022 and DE-AC02-98CH10886. We thank 
C. Bender for pointing out the concept of Chebyshev ellipses and W. Jancke for providing unpublished figures of double-peak distributions in a case of second order phase transition. The calculations were supported in part by a project of level C on the 
Fermilab cluster. \end{acknowledgments}

\appendix

\section{Multicanonical data}\label{app_muca}

\begin{figure}
\includegraphics[width= 3.3in]{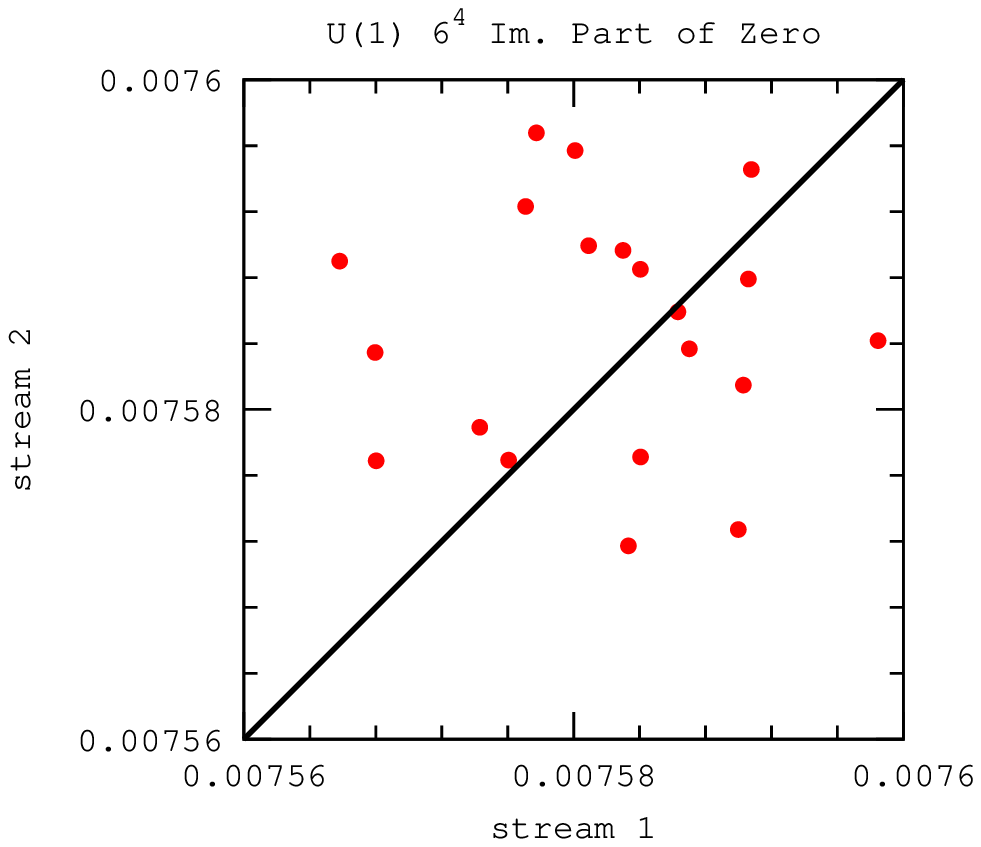}
\includegraphics[width= 3.3in]{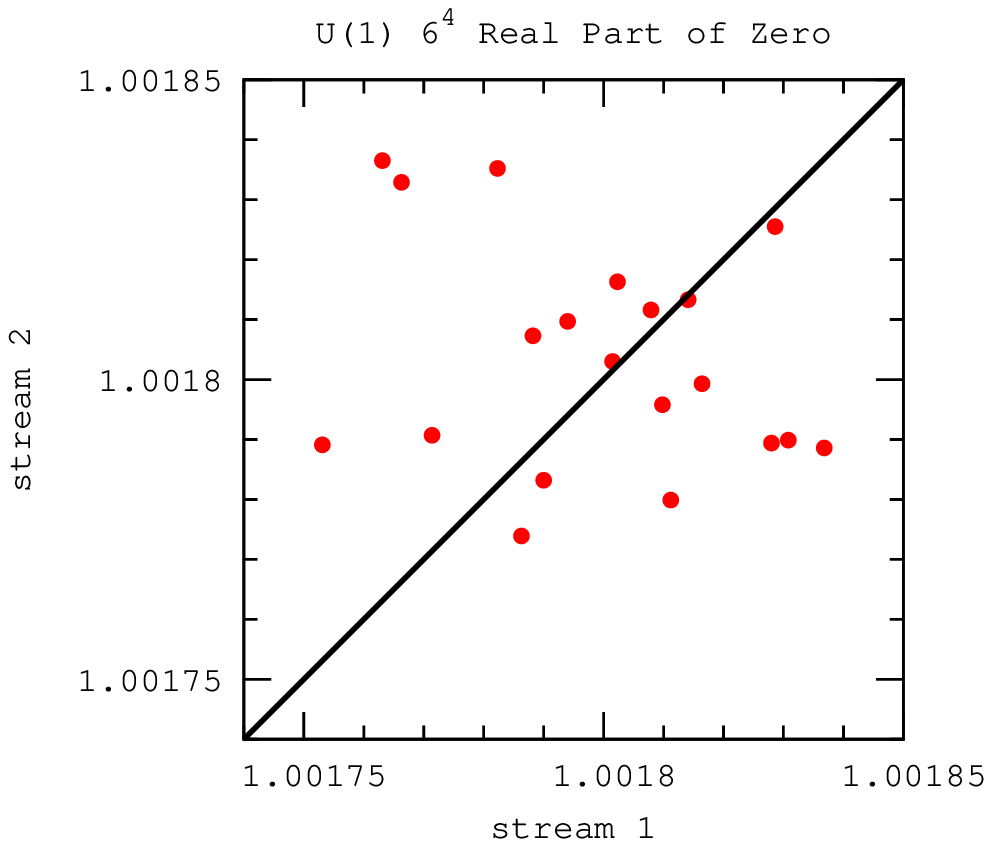}
\caption{\label{fig:RStream} Imaginary (top) and Real (bottom) part of the complex zeros of MUCA run 1 versus MUCA run 2 for the 20 streams on a $6^4$ lattice.}
\end{figure}

The number of tunnelings and the integrated autocorrelation times
for $4^4$, $6^4$ and $8^4$ lattices is given in
Tables~\ref{tab:muca_4}-\ref{tab:muca_8}. 
Fig. \ref{fig:RStream}  illustrates the fluctuations among streams and the correlations among MUCA runs for the zeros on a $6^4$ lattice. 
%We see that some streams have some MUCA runs very close to each other but far from the total average 
%and some other MUCA runs are about as far apart as they are from the total average. 
In table~\ref{tab:muca_large}
we summarize the parameters of simulations on $10^4$--$20^4$ lattices. 
%that were performed elsewhere~\cite{Berg:2006hh}, \cite{Baz_Berg_unpub}.

\begin{table}
\begin{center}
\begin{tabular}{||r|r|r|c|r|c||}
\hline
   & MUCA 1& \multicolumn{2}{|c|}{MUCA 2} & \multicolumn{2}{|c||}{MUCA 3}  \\\hline
\# & $N^{tunn}$ & $N^{tunn}$ & $\tau_{int}$ & $N^{tunn}$ & $\tau_{int}$ \\\hline
 1 & 1,512 & 3,886 &  99(2)       & 4,697 & 110(1)     \\
 2 & 1,406 & 4,018 & 109(1)       & 2,833 & 720(35)    \\
 3 & 1,974 & 4,723 & 114(5)       & 3,492 & 540(15)    \\
 4 & 1,552 & 4,537 & 216(5)       & 2,684 & 750(18)    \\
 5 &   774 & 5,038 & 175(12)      & 4,920 & 367(29)    \\
 6 &   963 & 4,769 & 109(1)       & 5,682 & 164(4)     \\
 7 &   196 &   397 & 1383(124)    & 3,162 & 679(34)    \\
 8 & 4,089 & 3,875 & 101(5)       & 4,547 & 116(5)     \\
 9 & 2,344 & 4,214 & 108(7)       & 4,599 & 266(6)     \\
10 & 1,652 & 4,582 & 185(15)      & 3,484 & 625(48)    \\
11 & 1,622 & 4,179 & 111(7)       & 5,030 & 255(12)    \\
12 & 1,722 & 4,281 & 126(3)       & 3,985 & 674(101)   \\
13 & 3,406 & 4,081 & 98(1)        & 4,994 & 146(3)     \\
14 & 1,271 & 4,127 & 104(6)       & 5,257 & 135(6)     \\
15 &   488 & 4,610 & 255(9)       & 3,776 & 524(14)    \\
16 & 2,351 & 4,394 & 108(3)       & 4,167 & 338(12)    \\
17 &   788 & 4,785 & 123(4)       & 4,598 & 356(10)    \\
18 &   735 & 4,680 & 134(4)       & 4,661 & 364(16)    \\
19 &   845 & 4,450 & 200(4)       & 3,675 & 537(14)    \\
20 & 2,697 & 3,526 &  93(1)       & 4,123 & 104(7)     \\
\hline
\end{tabular}
\end{center}
\caption{\label{tab:muca_4} MUCA data for $4^4$.}
\end{table}

\begin{table}
\begin{center}
\begin{tabular}{||r|r|c|r|c||}
\hline
   & \multicolumn{2}{|c|}{MUCA 1} & \multicolumn{2}{|c||}{MUCA 2}  \\\hline
\# & $N^{tunn}$ & $\tau_{int}$ & $N^{tunn}$ & $\tau_{int}$ \\\hline
 1 & 1,351 & 702(48)      &  950  & 559(27)    \\
 2 &   708 & 466(12)      &  897  & 480(11)    \\
 3 &   367 & 422(12)      &  947  & 502(17)    \\
 4 &   454 & 474(30)      &  971  & 561(31)    \\
 5 &   580 & 484(37)      &  894  & 565(47)    \\
 6 &   682 & 481(21)      &  911  & 511(23)    \\
 7 &   523 & 423(14)      &  909  & 557(43)    \\
 8 &   765 & 512(28)      &  903  & 532(33)    \\
 9 &   696 & 510(27)      &  921  & 485(12)    \\
10 &   652 & 469(32)      &  935  & 569(53)    \\
11 &   513 & 544(46)      &  976  & 523(33)    \\
12 &   378 & 396(18)      &  976  & 536(27)    \\
13 &   867 & 559(17)      &  961  & 495(10)    \\
14 &   661 & 496(13)      &  932  & 509(28)    \\
15 &   545 & 542(50)      &  896  & 554(38)    \\
16 &   615 & 497(14)      &  920  & 496(13)    \\
17 &   475 & 438(16)      &  979  & 543(46)    \\
18 &   822 & 464(10)      &  903  & 486(11)    \\
19 &   878 & 570(35)      &  892  & 508(30)    \\
20 &   578 & 588(62)      &  949  & 572(28)    \\
\hline
\end{tabular}
\end{center}
\caption{\label{tab:muca_6} MUCA data for $6^4$.}
\end{table}

\begin{table}
\begin{center}
\begin{tabular}{||r|r|c|r|c||}
\hline
   & \multicolumn{2}{|c|}{MUCA 1} & \multicolumn{2}{|c||}{MUCA 2}  \\\hline
\# & $N^{tunn}$ & $\tau_{int}$ & $N^{tunn}$ & $\tau_{int}$ \\\hline
 1 &  145  & 1,756(99)    &  252  & 1,782(124) \\
 2 &   79  & 1,298(69)    &  240  & 1,798(78)  \\
 3 &   75  & 1,475(144)   &  259  & 2,486(368) \\
 4 &  121  & 1,291(141)   &  216  & 2,159(235) \\
 5 &  150  & 2,420(364)   &  216  & 1,660(154) \\
 6 &   86  & 1,097(40)    &  256  & 1,528(55)  \\
 7 &   74  & 1,103(51)    &  255  & 1,621(105) \\
 8 &  132  & 1,419(48)    &  254  & 1,568(98)  \\
 9 &  187  & 3,089(438)   &  197  & 3,074(397) \\
10 &   98  & 1,370(94)    &  255  & 1,706(104) \\
11 &  142  & 1,389(47)    &  208  & 1,700(218) \\
12 &  165  & 1,804(343)   &  254  & 1,800(256) \\
13 &   93  & 1,265(111)   &  270  & 1,912(160) \\
14 &  212  & 2,012(114)   &  231  & 1,855(254) \\
15 &  137  & 1,581(181)   &  249  & 1,855(170) \\
16 &  159  & 1,904(235)   &  211  & 1,674(224) \\
17 &  269  & 1,773(76)    &  213  & 2,169(256) \\
18 &  206  & 1,773(176)   &  234  & 1,503(52)  \\
19 &  214  & 1,756(76)    &  212  & 1,954(265) \\
20 &   96  & 1,680(221)   &  227  & 1,697(101) \\
\hline
\end{tabular}
\end{center}
\caption{\label{tab:muca_8} MUCA data for $8^4$.}
\end{table}

\begin{table}
\begin{center}
\begin{tabular}{||r|r|r|r|r|r||}
\hline
volume & sweeps            & $\beta_{min}$ & $\beta_{max}$ & MUCA1 & MUCA2 \\
\hline
$10^4$ & $32\times 96,000$ & 0.980         & 1.030         & 103  & 133 \\
$12^4$ & $32\times112,000$ & 0.990         & 1.030         &  75  &  82 \\
$14^4$ & $32\times128,000$ & 1.000         & 1.020         &  57  &  51 \\
$20^4$ & $64\times100,000$ & 1.010         & 1.012         & 155  & 210 \\
\hline
\end{tabular}
\caption{\label{tab:muca_large} MUCA data for volumes where simulations
were performed in narrow $[\beta_{min},\beta_{max}]$ range. Last two 
columns summarize the number of tunneling events. $10^4$ -- $12^4$. is
from~\cite{Berg:2006hh}.}
\end{center}
\end{table}

\section{Numerical data used in Sec. \ref{sec:largeV}}
\label{sec:num}

In order to calculate the quantities used to make the  graphs of Sec. \ref{sec:largeV}, we have first fitted $f(s)$ in a region slightly wider than the location of the two peaks of $f(s)-\beta_S$ with the first 12 Chebyshev polynomials. Higher order polynomials tend to pick up the noise and provide results which are less stable. Using this polynomial approximation, we calculated the two roots of $f''(s)$ in the interval considered. We call them $s_L$ (Left) and $s_R$ (Right). They are the local extrema of $f'(s)$. The corresponding $\beta$ (obtained from the saddle point Eq. (\ref{eq:saddle}) are denoted $\beta_L$ and $\beta_R$. 
%$The height of kink is $\beta_R-\beta_L$. 
For $\beta_L<\beta<\beta_R$, the saddle point Eq. (\ref{eq:saddle}) has three solutions instead of one (the ``Maxwell kink"). $\beta_S$ corresponds to the case where the area of the kink below and above are equal. The location of the maxima of $f(s)-\beta_S s$ are called $s_1$ and $s_2$ as in Sec. \ref{sec:crossover}. $\Delta( f-\beta_S s)$  denotes the difference between the local minimum and the local maxima of $ f-\beta_S s$. The numerical results are provided in Table \ref{tab:scaling}.  
\newpage
%\begin{array}{cccccccccc}
\begin{widetext}
\begin{center}
\begin{table}
\begin{tabular}{||c|c|c|c|c|c|c|c|c|c||}
\hline
$L$ & $\beta_S$ & $s_1$ & $s_2$  &
$\Delta( f-\beta_S s)$ &  $s_L$ & $s_R$ & $\beta_L$ & $\beta_R$ & $\beta_R-\beta_L$
 \cr
\hline
 4 & 0.979327 & 0.369215 & 0.442409 & 0.0000597527 & 0.384952 & 0.425551 & 0.976759 &
   0.981792 & 0.00503262  \cr
 6 & 1.00176 & 0.352413 & 0.410538 & 0.0000533197 & 0.364762 & 0.395541 & 0.998768 &
   1.00446 & 0.00569344  \cr
 8 & 1.00738 & 0.348089 & 0.394932 & 0.0000337064 & 0.358049 & 0.382876 & 1.00504 &
   1.00951 & 0.00447297  \cr
 10 & 1.00942 & 0.345567 & 0.386254 & 0.000023194 & 0.353401 & 0.37509 & 1.00747 &
   1.01103 & 0.00356233  \cr
 12 & 1.01022 & 0.345633 & 0.380806 & 0.0000164495 & 0.352382 & 0.370685 & 1.0086 &
   1.01153 & 0.00293064  \cr
 14 & 1.01058 & 0.345352 & 0.377624 & 0.0000114541 & 0.351326 & 0.368136 & 1.00934 &
   1.01155 & 0.00221252  \cr
 20 & 1.01097 & 0.345337 & 0.374469 & 7.1542782 $10^{-6}$ &
   0.349468 & 0.36223 & 1.00988 & 1.01157 & 0.0016911 \cr
\hline
\end{tabular}
\caption{ \label{tab:scaling} 
 Numerical results described in the text. }
\end{table}
\end{center}
 \end{widetext}
 \newpage
%\bibliography{macbibF}

\begin{thebibliography}{10}%
\makeatletter
\providecommand \@ifxundefined [1]{%
 \ifx #1\undefined \expandafter \@firstoftwo
 \else \expandafter \@secondoftwo
\fi
}%
\providecommand \@ifnum [1]{%
 \ifnum #1\expandafter \@firstoftwo
 \else \expandafter \@secondoftwo
\fi
}%
\providecommand \enquote [1]{``#1''}%
\providecommand \bibnamefont  [1]{#1}%
\providecommand \bibfnamefont [1]{#1}%
\providecommand \citenamefont [1]{#1}%
\providecommand\href[0]{\@sanitize\@href}%
\providecommand\@href[1]{\endgroup\@@startlink{#1}\endgroup\@@href}%
\providecommand\@@href[1]{#1\@@endlink}%
\providecommand \@sanitize [0]{\begingroup\catcode`\&12\catcode`\#12\relax}%
\@ifxundefined \pdfoutput {\@firstoftwo}{%
 \@ifnum{\z@=\pdfoutput}{\@firstoftwo}{\@secondoftwo}%
}{%
 \providecommand\@@startlink[1]{\leavevmode}%
 \providecommand\@@endlink[0]{}%
}{%
 \providecommand\@@startlink[1]{%
  \leavevmode
  \pdfstartlink
   attr{/Border[0 0 1 ]/H/I/C[0 1 1]}%
   user{/Subtype/Link/A<</Type/Action/S/URI/URI(#1)>>}%
  \relax
 }%
 \providecommand\@@endlink[0]{\pdfendlink}%
}%
\providecommand \url  [0]{\begingroup\@sanitize \@url }%
\providecommand \@url [1]{\endgroup\@href {#1}{\urlprefix}}%
\providecommand \urlprefix [0]{URL }%
\providecommand \Eprint[0]{\href }%
\@ifxundefined \urlstyle {%
  \providecommand \doi [1]{doi:\discretionary{}{}{}#1}%
}{%
  \providecommand \doi [0]{doi:\discretionary{}{}{}\begingroup
  \urlstyle{rm}\Url }%
}%
\providecommand \doibase [0]{http://dx.doi.org/}%
\providecommand \Doi[1]{\href{\doibase#1}}%
\providecommand \bibAnnote [3]{%
  \BibitemShut{#1}%
  \begin{quotation}\noindent
    \textsc{Key:}\ #2\\\textsc{Annotation:}\ #3%
  \end{quotation}%
}%
\providecommand \bibAnnoteFile [2]{%
  \IfFileExists{#2}{\bibAnnote {#1} {#2} {\input{#2}}}{}%
}%
\providecommand \typeout [0]{\immediate \write \m@ne }%
\providecommand \selectlanguage [0]{\@gobble}%
\providecommand \bibinfo [0]{\@secondoftwo}%
\providecommand \bibfield [0]{\@secondoftwo}%
\providecommand \translation [1]{[#1]}%
\providecommand \BibitemOpen[0]{}%
\providecommand \bibitemStop [0]{}%
\providecommand \bibitemNoStop [0]{.\EOS\space}%
\providecommand \EOS [0]{\spacefactor3000\relax}%
\providecommand \BibitemShut [1]{\csname bibitem#1\endcsname}%
%</preamble>
\bibitem{appelquist09}%
  \BibitemOpen
  \bibfield{author}{%
  \bibinfo {author} {\bibfnamefont{T.}~\bibnamefont{Appelquist}}, \bibinfo
  {author} {\bibfnamefont{G.~T.}\ \bibnamefont{Fleming}},\ and\ \bibinfo
  {author} {\bibfnamefont{E.~T.}\ \bibnamefont{Neil}},\ }%
  \bibfield{journal}{%
  \Doi{10.1103/PhysRevD.79.076010}{\bibinfo {journal} {Phys. Rev. D}}\ }%
  \textbf{\bibinfo {volume} {79}},\ \bibinfo {pages} {076010} (\bibinfo {year}
  {2009}),\ \Eprint{http://arxiv.org/abs/0901.3766}{arXiv:0901.3766 [hep-ph]}%
  \bibAnnoteFile{NoStop}{appelquist09}%
%%CITATION = 0901.3766;%%
\bibitem{hasenfratz09}%
  \BibitemOpen
  \bibfield{author}{%
  \bibinfo {author} {\bibfnamefont{A.}~\bibnamefont{Hasenfratz}},\ }%
  \bibfield{journal}{%
  \Doi{10.1103/PhysRevD.80.034505}{\bibinfo {journal} {Phys. Rev. D}}\ }%
  \textbf{\bibinfo {volume} {80}},\ \bibinfo {pages} {034505} (\bibinfo {year}
  {2009}),\ \Eprint{http://arxiv.org/abs/0907.0919}{arXiv:0907.0919 [hep-lat]}%
  \bibAnnoteFile{NoStop}{hasenfratz09}%
%%CITATION = 0907.0919;%%
\bibitem{fodor09}%
  \BibitemOpen
  \bibfield{author}{%
  \bibinfo {author} {\bibfnamefont{Z.}~\bibnamefont{Fodor}}, \bibinfo {author}
  {\bibfnamefont{K.}~\bibnamefont{Holland}}, \bibinfo {author}
  {\bibfnamefont{J.}~\bibnamefont{Kuti}}, \bibinfo {author}
  {\bibfnamefont{D.}~\bibnamefont{Nogradi}},\ and\ \bibinfo {author}
  {\bibfnamefont{C.}~\bibnamefont{Schroeder}},\ }%
  \bibfield{journal}{%
  \Doi{10.1016/j.physletb.2009.10.040}{\bibinfo {journal} {Phys. Lett. B}}\ }%
  \textbf{\bibinfo {volume} {681}},\ \bibinfo {pages} {353} (\bibinfo {year}
  {2009}),\ \Eprint{http://arxiv.org/abs/0907.4562}{arXiv:0907.4562 [hep-lat]}%
  \bibAnnoteFile{NoStop}{fodor09}%
%%CITATION = 0907.4562;%%
\bibitem{Deuzeman:2009mh}%
  \BibitemOpen
  \bibfield{author}{%
  \bibinfo {author} {\bibfnamefont{A.}~\bibnamefont{Deuzeman}}, \bibinfo
  {author} {\bibfnamefont{M.~P.}\ \bibnamefont{Lombardo}},\ and\ \bibinfo
  {author} {\bibfnamefont{E.}~\bibnamefont{Pallante}},\ }%
  \bibfield{journal}{%
  \Doi{10.1103/PhysRevD.82.074503}{\bibinfo {journal} {Phys. Rev. D}}\ }%
  \textbf{\bibinfo {volume} {82}},\ \bibinfo {pages} {074503} (\bibinfo {year}
  {2010}),\ \Eprint{http://arxiv.org/abs/0904.4662}{arXiv:0904.4662 [hep-ph]}%
  \bibAnnoteFile{NoStop}{Deuzeman:2009mh}%
%%CITATION = 0904.4662;%%
\bibitem{Fodor:2011tu}%
  \BibitemOpen
  \bibfield{author}{%
  \bibinfo {author} {\bibfnamefont{Z.}~\bibnamefont{Fodor}}, \bibinfo {author}
  {\bibfnamefont{K.}~\bibnamefont{Holland}}, \bibinfo {author}
  {\bibfnamefont{J.}~\bibnamefont{Kuti}}, \bibinfo {author}
  {\bibfnamefont{D.}~\bibnamefont{Nogradi}},\ and\ \bibinfo {author}
  {\bibfnamefont{C.}~\bibnamefont{Schroeder}},\ }%
  \bibfield{journal}{%
  \Doi{10.1016/j.physletb.2011.07.037}{\bibinfo {journal} {Phys. Lett. B}}\ }%
  \textbf{\bibinfo {volume} {703}},\ \bibinfo {pages} {348} (\bibinfo {year}
  {2011}),\ \Eprint{http://arxiv.org/abs/1104.3124}{arXiv:1104.3124 [hep-lat]}%
  \bibAnnoteFile{NoStop}{Fodor:2011tu}%
\bibitem{Hasenfratz:2010fi}%
  \BibitemOpen
  \bibfield{author}{%
  \bibinfo {author} {\bibfnamefont{A.}~\bibnamefont{Hasenfratz}},\ }%
  \bibfield{journal}{%
  \Doi{10.1103/PhysRevD.82.014506}{\bibinfo {journal} {Phys. Rev. D}}\ }%
  \textbf{\bibinfo {volume} {82}},\ \bibinfo {pages} {014506} (\bibinfo {year}
  {2010}),\ \Eprint{http://arxiv.org/abs/1004.1004}{arXiv:1004.1004 [hep-lat]}%
  \bibAnnoteFile{NoStop}{Hasenfratz:2010fi}%
\bibitem{shamir08}%
  \BibitemOpen
  \bibfield{author}{%
  \bibinfo {author} {\bibfnamefont{Y.}~\bibnamefont{Shamir}}, \bibinfo {author}
  {\bibfnamefont{B.}~\bibnamefont{Svetitsky}},\ and\ \bibinfo {author}
  {\bibfnamefont{T.}~\bibnamefont{DeGrand}},\ }%
  \bibfield{journal}{%
  \Doi{10.1103/PhysRevD.78.031502}{\bibinfo {journal} {Phys. Rev. D}}\ }%
  \textbf{\bibinfo {volume} {78}},\ \bibinfo {pages} {031502} (\bibinfo {year}
  {2008}),\ \Eprint{http://arxiv.org/abs/0803.1707}{arXiv:0803.1707 [hep-lat]}%
  \bibAnnoteFile{NoStop}{shamir08}%
%%CITATION = 0803.1707;%%
\bibitem{2010PhRvD..81k4507K}%
  \BibitemOpen
  \bibfield{author}{%
  \bibinfo {author} {\bibfnamefont{J.~B.}\ \bibnamefont{{Kogut}}}\ and\
  \bibinfo {author} {\bibfnamefont{D.~K.}\ \bibnamefont{{Sinclair}}},\ }%
  \bibfield{journal}{%
  \Doi{10.1103/PhysRevD.81.114507}{\bibinfo {journal} {\prd}}\ }%
  \textbf{\bibinfo {volume} {81}},\ \bibinfo {pages} {114507} (\bibinfo {month}
  {Jun.}\ \bibinfo {year} {2010}),\
  \Eprint{http://arxiv.org/abs/1002.2988}{arXiv:1002.2988 [hep-lat]}%
  \bibAnnoteFile{NoStop}{2010PhRvD..81k4507K}%
\bibitem{DeGrand:2011qd}%
  \BibitemOpen
  \bibfield{author}{%
  \bibinfo {author} {\bibfnamefont{T.}~\bibnamefont{DeGrand}}, \bibinfo
  {author} {\bibfnamefont{Y.}~\bibnamefont{Shamir}},\ and\ \bibinfo {author}
  {\bibfnamefont{B.}~\bibnamefont{Svetitsky}},\ }%
  \bibfield{journal}{%
  \Doi{10.1103/PhysRevD.83.074507}{\bibinfo {journal} {Phys. Rev. D}}\ }%
  \textbf{\bibinfo {volume} {83}},\ \bibinfo {pages} {074507} (\bibinfo {year}
  {2011}),\ \Eprint{http://arxiv.org/abs/1102.2843}{arXiv:1102.2843 [hep-lat]}%
  \bibAnnoteFile{NoStop}{DeGrand:2011qd}%
\bibitem{Sinclair:2010be}%
  \BibitemOpen
  \bibfield{author}{%
  \bibinfo {author} {\bibfnamefont{D.~K.}\ \bibnamefont{Sinclair}}\ and\
  \bibinfo {author} {\bibfnamefont{J.~.~B.}\ \bibnamefont{Kogut}},\ }%
  \bibfield{journal}{%
  \bibinfo {journal} {PoS}\ }%
  \textbf{\bibinfo {volume} {LATTICE2010}},\ \bibinfo {pages} {071} (\bibinfo
  {year} {2010}),\ \Eprint{http://arxiv.org/abs/1008.2468}{arXiv:1008.2468 [hep-lat]}%
  \bibAnnoteFile{NoStop}{Sinclair:2010be}%
\bibitem{Kogut:2011ty}%
  \BibitemOpen
  \bibfield{author}{%
  \bibinfo {author} {\bibfnamefont{J.~B.}\ \bibnamefont{Kogut}}\ and\ \bibinfo
  {author} {\bibfnamefont{D.~K.}\ \bibnamefont{Sinclair}}}%
   (\bibinfo {year} {2011}),\
  \Eprint{http://arxiv.org/abs/1105.3749}{arXiv:1105.3749 [hep-lat]}%
  \bibAnnoteFile{NoStop}{Kogut:2011ty}%
%%CITATION = 1105.3749;%%
\bibitem{DelDebbio:2010ze}%
  \BibitemOpen
  \bibfield{author}{%
  \bibinfo {author} {\bibfnamefont{L.}~\bibnamefont{Del~Debbio}}\ and\ \bibinfo
  {author} {\bibfnamefont{R.}~\bibnamefont{Zwicky}},\ }%
  \bibfield{journal}{%
  \Doi{10.1103/PhysRevD.82.014502}{\bibinfo {journal} {Phys. Rev. D}}\ }%
  \textbf{\bibinfo {volume} {82}},\ \bibinfo {pages} {014502} (\bibinfo {year}
  {2010}),\ \Eprint{http://arxiv.org/abs/1005.2371}{arXiv:1005.2371 [hep-ph]}%
  \bibAnnoteFile{NoStop}{DelDebbio:2010ze}%
%%CITATION = 1005.2371;%%
\bibitem{Myers:2009df}%
  \BibitemOpen
  \bibfield{author}{%
  \bibinfo {author} {\bibfnamefont{J.~C.}\ \bibnamefont{Myers}}\ and\ \bibinfo
  {author} {\bibfnamefont{M.~C.}\ \bibnamefont{Ogilvie}},\ }%
  \bibfield{journal}{%
  \Doi{10.1088/1126-6708/2009/07/095}{\bibinfo {journal} {JHEP}}\ }%
  \textbf{\bibinfo {volume} {07}},\ \bibinfo {pages} {095} (\bibinfo {year}
  {2009}),\ \Eprint{http://arxiv.org/abs/0903.4638}{arXiv:0903.4638 [hep-th]}%
  \bibAnnoteFile{NoStop}{Myers:2009df}%
%%CITATION = 0903.4638;%%
\bibitem{DeGrand:2010ba}%
  \BibitemOpen
  \bibfield{author}{%
  \bibinfo {author} {\bibfnamefont{T.}~\bibnamefont{DeGrand}},\ }%
  \bibfield{journal}{%
  \bibinfo {journal} {Phil. Trans. R. Soc. A}\ }%
  \textbf{\bibinfo {volume} {369}},\ \bibinfo {pages} {2701} (\bibinfo {year}
  {2011}),\ \Eprint{http://arxiv.org/abs/1010.4741}{arXiv:1010.4741 [hep-lat]}%
  \bibAnnoteFile{NoStop}{DeGrand:2010ba}%
%%CITATION = 1010.4741;%%
\bibitem{Ogilvie:2010vx}%
  \BibitemOpen
  \bibfield{author}{%
  \bibinfo {author} {\bibfnamefont{M.~C.}\ \bibnamefont{Ogilvie}},\ }%
  \bibfield{journal}{%
  \bibinfo {journal} {Phil. Trans. R. Soc. A}\ }%
  \textbf{\bibinfo {volume} {369}},\ \bibinfo {pages} {2718} (\bibinfo {year}
  {2011}),\ \Eprint{http://arxiv.org/abs/1010.1942}{arXiv:1010.1942 [hep-lat]}%
  \bibAnnoteFile{NoStop}{Ogilvie:2010vx}%
%%CITATION = 1010.1942;%%
\bibitem{Sannino:2009za}%
  \BibitemOpen
  \bibfield{author}{%
  \bibinfo {author} {\bibfnamefont{F.}~\bibnamefont{Sannino}},\ }%
  \bibfield{journal}{%
  \bibinfo {journal} {Acta Phys. Polon. B}\ }%
  \textbf{\bibinfo {volume} {40}},\ \bibinfo {pages} {3533} (\bibinfo {year}
  {2009}),\ \Eprint{http://arxiv.org/abs/0911.0931}{arXiv:0911.0931 [hep-ph]}%
  \bibAnnoteFile{NoStop}{Sannino:2009za}%
\bibitem{Berg:2010hj}%
  \BibitemOpen
  \bibfield{author}{%
  \bibinfo {author} {\bibfnamefont{B.~A.}\ \bibnamefont{Berg}},\ }%
  \bibfield{journal}{%
  \Doi{10.1103/PhysRevD.82.114507}{\bibinfo {journal} {Phys. Rev.}}\ }%
  \textbf{\bibinfo {volume} {D82}},\ \bibinfo {pages} {114507} (\bibinfo {year}
  {2010}),\ \Eprint{http://arxiv.org/abs/1011.6406}{arXiv:1011.6406 [hep-lat]}%
  \bibAnnoteFile{NoStop}{Berg:2010hj}%
%%CITATION = 1011.6406;%%
\bibitem{Berg:2011sz}%
  \BibitemOpen
  \bibfield{author}{%
  \bibinfo {author} {\bibfnamefont{B.~A.}\ \bibnamefont{Berg}}}%
   (\bibinfo {year} {2011}),\
  \Eprint{http://arxiv.org/abs/1109.5861}{arXiv:1109.5861 [hep-lat]}%
  \bibAnnoteFile{NoStop}{Berg:2011sz}%
%%CITATION = 1109.5861;%%
\bibitem{Denbleyker:2010sv}%
  \BibitemOpen
  \bibfield{author}{%
  \bibinfo {author} {\bibfnamefont{A.}~\bibnamefont{Denbleyker}}, \bibinfo
  {author} {\bibfnamefont{D.}~\bibnamefont{Du}}, \bibinfo {author}
  {\bibfnamefont{Y.}~\bibnamefont{Liu}}, \bibinfo {author}
  {\bibfnamefont{Y.}~\bibnamefont{Meurice}},\ and\ \bibinfo {author}
  {\bibfnamefont{H.}~\bibnamefont{Zou}},\ }%
  \bibfield{journal}{%
  \Doi{10.1103/PhysRevLett.104.251601}{\bibinfo {journal} {Phys. Rev. Lett.}}\
  }%
  \textbf{\bibinfo {volume} {104}},\ \bibinfo {pages} {251601} (\bibinfo {year}
  {2010}),\ \Eprint{http://arxiv.org/abs/1005.1993}{arXiv:1005.1993 [hep-lat]}%
  \bibAnnoteFile{NoStop}{Denbleyker:2010sv}%
%%CITATION = 1005.1993;%%
\bibitem{PhysRevD.83.056009}%
  \BibitemOpen
  \bibfield{author}{%
  \bibinfo {author} {\bibfnamefont{Y.}~\bibnamefont{{Meurice}}}\ and\ \bibinfo
  {author} {\bibfnamefont{H.}~\bibnamefont{{Zou}}},\ }%
  \bibfield{journal}{%
  \Doi{10.1103/PhysRevD.83.056009}{\bibinfo {journal} {\prd}}\ }%
  \textbf{\bibinfo {volume} {83}},\ \bibinfo {pages} {056009} (\bibinfo {month}
  {Mar.}\ \bibinfo {year} {2011}),\
  \Eprint{http://arxiv.org/abs/1101.1319}{arXiv:1101.1319 [hep-lat]}%
  \bibAnnoteFile{NoStop}{PhysRevD.83.056009}%
\bibitem{Liu:2011zzh}%
  \BibitemOpen
  \bibfield{author}{%
  \bibinfo {author} {\bibfnamefont{Y.}~\bibnamefont{Liu}}\ and\ \bibinfo
  {author} {\bibfnamefont{Y.}~\bibnamefont{Meurice}},\ }%
  \bibfield{journal}{%
  \Doi{10.1103/PhysRevD.83.096008}{\bibinfo {journal} {Phys. Rev. D}}\ }%
  \textbf{\bibinfo {volume} {83}},\ \bibinfo {pages} {096008} (\bibinfo {year}
  {2011}),\ \Eprint{http://arxiv.org/abs/1103.4846}{arXiv:1103.4846 [hep-lat]}%
  \bibAnnoteFile{NoStop}{Liu:2011zzh}%
\bibitem{Tomboulis:2009zz}%
  \BibitemOpen
  \bibfield{author}{%
  \bibinfo {author} {\bibfnamefont{E.~T.}\ \bibnamefont{Tomboulis}},\ }%
  \bibfield{journal}{%
  \Doi{10.1142/S0217732309032307}{\bibinfo {journal} {Mod. Phys. Lett. A}}\ }%
  \textbf{\bibinfo {volume} {24}},\ \bibinfo {pages} {2717} (\bibinfo {year}
  {2009})%
  \bibAnnoteFile{NoStop}{Tomboulis:2009zz}%
%%CITATION = MPLAE,A24,2717;%%
\bibitem{fisher65}%
  \BibitemOpen
  \bibfield{author}{%
  \bibinfo {author} {\bibfnamefont{M.}~\bibnamefont{Fisher}},\ }%
  \enquote{\bibinfo {title} {in lectures in theoretical physics vol. viic},}\ \
  (\bibinfo {publisher} {University of Colorado Press},\ \bibinfo {address}
  {Boulder, Colorado},\ \bibinfo {year} {1965})%
  \bibAnnoteFile{NoStop}{fisher65}%
\bibitem{alves90b}%
  \BibitemOpen
  \bibfield{author}{%
  \bibinfo {author} {\bibfnamefont{N.~A.}\ \bibnamefont{Alves}}, \bibinfo
  {author} {\bibfnamefont{B.~A.}\ \bibnamefont{Berg}},\ and\ \bibinfo {author}
  {\bibfnamefont{S.}~\bibnamefont{Sanielevici}},\ }%
  \bibfield{journal}{%
  \bibinfo {journal} {Phys. Rev. Lett.}\ }%
  \textbf{\bibinfo {volume} {64}},\ \bibinfo {pages} {3107} (\bibinfo {year}
  {1990})%
  \bibAnnoteFile{NoStop}{alves90b}%
%%CITATION = PRLTA,64,3107;%%
\bibitem{Jersak:1996mn}%
  \BibitemOpen
  \bibfield{author}{%
  \bibinfo {author} {\bibfnamefont{J.}~\bibnamefont{Jersak}}, \bibinfo {author}
  {\bibfnamefont{C.~B.}\ \bibnamefont{Lang}},\ and\ \bibinfo {author}
  {\bibfnamefont{T.}~\bibnamefont{Neuhaus}},\ }%
  \bibfield{journal}{%
  \Doi{10.1103/PhysRevLett.77.1933}{\bibinfo {journal} {Phys. Rev. Lett.}}\ }%
  \textbf{\bibinfo {volume} {77}},\ \bibinfo {pages} {1933} (\bibinfo {year}
  {1996}),\
  \Eprint{http://arxiv.org/abs/hep-lat/9606010}{arXiv:hep-lat/9606010}%
  \bibAnnoteFile{NoStop}{Jersak:1996mn}%
%%CITATION = HEP-LAT/9606010;%%
\bibitem{Jersak:1996mj}%
  \BibitemOpen
  \bibfield{author}{%
  \bibinfo {author} {\bibfnamefont{J.}~\bibnamefont{Jersak}}, \bibinfo {author}
  {\bibfnamefont{C.}~\bibnamefont{Lang}},\ and\ \bibinfo {author}
  {\bibfnamefont{T.}~\bibnamefont{Neuhaus}},\ }%
  \bibfield{journal}{%
  \Doi{10.1103/PhysRevD.54.6909}{\bibinfo {journal} {Phys.Rev. D}}\ }%
  \textbf{\bibinfo {volume} {54}},\ \bibinfo {pages} {6909} (\bibinfo {year}
  {1996}),\ \Eprint{http://arxiv.org/abs/hep-lat/9606013}{arXiv:hep-lat/9606013 [hep-lat]}%
  \bibAnnoteFile{NoStop}{Jersak:1996mj}%
\bibitem{janke00}%
  \BibitemOpen
  \bibfield{author}{%
  \bibinfo {author} {\bibfnamefont{W.}~\bibnamefont{Janke}}\ and\ \bibinfo
  {author} {\bibfnamefont{R.}~\bibnamefont{Kenna}},\ }%
  \bibfield{journal}{%
  \bibinfo {journal} {J. Stat. Phys.}\ }%
  \textbf{\bibinfo {volume} {102}},\ \bibinfo {pages} {1211} (\bibinfo {year}
  {2001}),\ \Eprint{http://arxiv.org/abs/cond-mat/0012026}{cond-mat/0012026}%
  \bibAnnoteFile{NoStop}{janke00}%
%%CITATION = COND-MAT/0012026;%%
\bibitem{janke01}%
  \BibitemOpen
  \bibfield{author}{%
  \bibinfo {author} {\bibfnamefont{W.}~\bibnamefont{Janke}}\ and\ \bibinfo
  {author} {\bibfnamefont{R.}~\bibnamefont{Kenna}},\ }%
  \bibfield{journal}{%
  \bibinfo {journal} {Nucl. Phys. Proc. Suppl.}\ }%
  \textbf{\bibinfo {volume} {106}},\ \bibinfo {pages} {905} (\bibinfo {year}
  {2002}),\ \Eprint{http://arxiv.org/abs/hep-lat/0112032}{hep-lat/0112032}%
  \bibAnnoteFile{NoStop}{janke01}%
%%CITATION = HEP-LAT 0112032;%%
\bibitem{janke04}%
  \BibitemOpen
  \bibfield{author}{%
  \bibinfo {author} {\bibfnamefont{W.}~\bibnamefont{Janke}}, \bibinfo {author}
  {\bibfnamefont{D.~A.}\ \bibnamefont{Johnston}},\ and\ \bibinfo {author}
  {\bibfnamefont{R.}~\bibnamefont{Kenna}},\ }%
  \bibfield{journal}{%
  \bibinfo {journal} {Nucl. Phys.}\ }%
  \textbf{\bibinfo {volume} {B682}},\ \bibinfo {pages} {618} (\bibinfo {year}
  {2004}),\ \Eprint{http://arxiv.org/abs/cond-mat/0401097}{cond-mat/0401097}%
  \bibAnnoteFile{NoStop}{janke04}%
%%CITATION = COND-MAT/0401097;%%
\bibitem{Klaus:1997be}%
  \BibitemOpen
  \bibfield{author}{%
  \bibinfo {author} {\bibfnamefont{B.}~\bibnamefont{Klaus}}\ and\ \bibinfo
  {author} {\bibfnamefont{C.}~\bibnamefont{Roiesnel}},\ }%
  \bibfield{journal}{%
  \Doi{10.1103/PhysRevD.58.114509}{\bibinfo {journal} {Phys. Rev. D}}\ }%
  \textbf{\bibinfo {volume} {58}},\ \bibinfo {pages} {114509} (\bibinfo {year}
  {1998}),\
  \Eprint{http://arxiv.org/abs/hep-lat/9801036}{arXiv:hep-lat/9801036}%
  \bibAnnoteFile{NoStop}{Klaus:1997be}%
%%CITATION = HEP-LAT/9801036;%%
\bibitem{Campos:1998jp}%
  \BibitemOpen
  \bibfield{author}{%
  \bibinfo {author} {\bibfnamefont{I.}~\bibnamefont{Campos}}, \bibinfo {author}
  {\bibfnamefont{A.}~\bibnamefont{Cruz}},\ and\ \bibinfo {author}
  {\bibfnamefont{A.}~\bibnamefont{Tarancon}},\ }%
  \bibfield{journal}{%
  \Doi{10.1016/S0550-3213(98)00452-0}{\bibinfo {journal} {Nucl.Phys.}}\ }%
  \textbf{\bibinfo {volume} {B528}},\ \bibinfo {pages} {325} (\bibinfo {year}
  {1998}),\ \Eprint{http://arxiv.org/abs/hep-lat/9803007}{arXiv:hep-lat/9803007
  [hep-lat]}%
  \bibAnnoteFile{NoStop}{Campos:1998jp}%
\bibitem{Arnold:2000hf}%
  \BibitemOpen
  \bibfield{author}{%
  \bibinfo {author} {\bibfnamefont{G.}~\bibnamefont{Arnold}}, \bibinfo {author}
  {\bibfnamefont{T.}~\bibnamefont{Lippert}}, \bibinfo {author}
  {\bibfnamefont{K.}~\bibnamefont{Schilling}},\ and\ \bibinfo {author}
  {\bibfnamefont{T.}~\bibnamefont{Neuhaus}},\ }%
  \bibfield{journal}{%
  \Doi{10.1016/S0920-5632(01)01001-5}{\bibinfo {journal}
  {Nucl.Phys.Proc.Suppl.}}\ }%
  \textbf{\bibinfo {volume} {94}},\ \bibinfo {pages} {651} (\bibinfo {year}
  {2001}),\ \Eprint{http://arxiv.org/abs/hep-lat/0011058}{arXiv:hep-lat/0011058 [hep-lat]}%
  \bibAnnoteFile{NoStop}{Arnold:2000hf}%
\bibitem{Arnold:2002jk}%
  \BibitemOpen
  \bibfield{author}{%
  \bibinfo {author} {\bibfnamefont{G.}~\bibnamefont{Arnold}}, \bibinfo {author}
  {\bibfnamefont{B.}~\bibnamefont{Bunk}}, \bibinfo {author}
  {\bibfnamefont{T.}~\bibnamefont{Lippert}},\ and\ \bibinfo {author}
  {\bibfnamefont{K.}~\bibnamefont{Schilling}},\ }%
  \bibfield{journal}{%
  \Doi{10.1016/S0920-5632(03)01704-3}{\bibinfo {journal} {Nucl. Phys. Proc.
  Suppl.}}\ }%
  \textbf{\bibinfo {volume} {119}},\ \bibinfo {pages} {864} (\bibinfo {year}
  {2003}),\
  \Eprint{http://arxiv.org/abs/hep-lat/0210010}{arXiv:hep-lat/0210010}%
  \bibAnnoteFile{NoStop}{Arnold:2002jk}%
%%CITATION = HEP-LAT/0210010;%%
\bibitem{Vettorazzo:2003fg}%
  \BibitemOpen
  \bibfield{author}{%
  \bibinfo {author} {\bibfnamefont{M.}~\bibnamefont{Vettorazzo}}\ and\ \bibinfo
  {author} {\bibfnamefont{P.}~\bibnamefont{de~Forcrand}},\ }%
  \bibfield{journal}{%
  \Doi{10.1016/j.nuclphysb.2004.02.038}{\bibinfo {journal} {Nucl.Phys.}}\ }%
  \textbf{\bibinfo {volume} {B686}},\ \bibinfo {pages} {85} (\bibinfo {year}
  {2004}),\ \Eprint{http://arxiv.org/abs/hep-lat/0311006}{arXiv:hep-lat/0311006 [hep-lat]}%
  \bibAnnoteFile{NoStop}{Vettorazzo:2003fg}%
\bibitem{BergNeuhaus1991}%
  \BibitemOpen
  \bibfield{author}{%
  \bibinfo {author} {\bibfnamefont{B.~A.}\ \bibnamefont{Berg}}\ and\ \bibinfo
  {author} {\bibfnamefont{T.}~\bibnamefont{Neuhaus}},\ }%
  \bibfield{journal}{%
  \bibinfo {journal} {Phys. Lett.}\ }%
  \textbf{\bibinfo {volume} {B267}},\ \bibinfo {pages} {249} (\bibinfo {year}
  {1991})%
  \bibAnnoteFile{NoStop}{BergNeuhaus1991}%
\bibitem{WangLandau2001}%
  \BibitemOpen
  \bibfield{author}{%
  \bibinfo {author} {\bibfnamefont{F.}~\bibnamefont{Wang}}\ and\ \bibinfo
  {author} {\bibfnamefont{D.~P.}\ \bibnamefont{Landau}},\ }%
  \bibfield{journal}{%
  \bibinfo {journal} {Phys. Rev. Lett.}\ }%
  \textbf{\bibinfo {volume} {86}} (\bibinfo {year} {2001})%
  \bibAnnoteFile{NoStop}{WangLandau2001}%
\bibitem{Bazavov:2005zy}%
  \BibitemOpen
  \bibfield{author}{%
  \bibinfo {author} {\bibfnamefont{A.}~\bibnamefont{Bazavov}}\ and\ \bibinfo
  {author} {\bibfnamefont{B.~A.}\ \bibnamefont{Berg}},\ }%
  \bibfield{journal}{%
  \Doi{10.1103/PhysRevD.71.114506}{\bibinfo {journal} {Phys. Rev. D}}\ }%
  \textbf{\bibinfo {volume} {71}},\ \bibinfo {pages} {114506} (\bibinfo {year}
  {2005}),\
  \Eprint{http://arxiv.org/abs/hep-lat/0503006}{arXiv:hep-lat/0503006}%
  \bibAnnoteFile{NoStop}{Bazavov:2005zy}%
%%CITATION = HEP-LAT/0503006;%%
\bibitem{Bazavov:2009pj}%
  \BibitemOpen
  \bibfield{author}{%
  \bibinfo {author} {\bibfnamefont{A.}~\bibnamefont{Bazavov}}\ and\ \bibinfo
  {author} {\bibfnamefont{B.~A.}\ \bibnamefont{Berg}},\ }%
  \bibfield{journal}{%
  \Doi{10.1016/j.cpc.2009.07.006}{\bibinfo {journal} {Comput. Phys. Commun.}}\
  }%
  \textbf{\bibinfo {volume} {180}},\ \bibinfo {pages} {2339} (\bibinfo {year}
  {2009}),\ \Eprint{http://arxiv.org/abs/0903.3984}{arXiv:0903.3984 [hep-lat]}%
  \bibAnnoteFile{NoStop}{Bazavov:2009pj}%
%%CITATION = 0903.3984;%%
\bibitem{Denbleyker:2007dy}%
  \BibitemOpen
  \bibfield{author}{%
  \bibinfo {author} {\bibfnamefont{A.}~\bibnamefont{Denbleyker}}, \bibinfo
  {author} {\bibfnamefont{D.}~\bibnamefont{Du}}, \bibinfo {author}
  {\bibfnamefont{Y.}~\bibnamefont{Meurice}},\ and\ \bibinfo {author}
  {\bibfnamefont{A.}~\bibnamefont{Velytsky}},\ }%
  \bibfield{journal}{%
  \bibinfo {journal} {PoS}\ }%
  \textbf{\bibinfo {volume} {LAT2007}},\ \bibinfo {pages} {269} (\bibinfo
  {year} {2007}),\ \Eprint{http://arxiv.org/abs/0710.5771}{arXiv:0710.5771 [hep-lat]}%
  \bibAnnoteFile{NoStop}{Denbleyker:2007dy}%
%%CITATION = 0710.5771;%%
\bibitem{alves89}%
  \BibitemOpen
  \bibfield{author}{%
  \bibinfo {author} {\bibfnamefont{N.~A.}\ \bibnamefont{Alves}}, \bibinfo
  {author} {\bibfnamefont{B.~A.}\ \bibnamefont{Berg}},\ and\ \bibinfo {author}
  {\bibfnamefont{R.}~\bibnamefont{Villanova}},\ }%
  \bibfield{journal}{%
  \bibinfo {journal} {Phys. Rev. B}\ }%
  \textbf{\bibinfo {volume} {41}},\ \bibinfo {pages} {383} (\bibinfo {year}
  {1990})%
  \bibAnnoteFile{NoStop}{alves89}%
%%CITATION = PHRVA,B41,383;%%
\bibitem{alves91}%
  \BibitemOpen
  \bibfield{author}{%
  \bibinfo {author} {\bibfnamefont{N.~A.}\ \bibnamefont{Alves}}, \bibinfo
  {author} {\bibfnamefont{B.~A.}\ \bibnamefont{Berg}},\ and\ \bibinfo {author}
  {\bibfnamefont{S.}~\bibnamefont{Sanielevici}},\ }%
  \bibfield{journal}{%
  \bibinfo {journal} {Nucl. Phys.}\ }%
  \textbf{\bibinfo {volume} {B376}},\ \bibinfo {pages} {218} (\bibinfo {year}
  {1992}),\ \Eprint{http://arxiv.org/abs/hep-lat/9107002}{hep-lat/9107002}%
  \bibAnnoteFile{NoStop}{alves91}%
%%CITATION = HEP-LAT 9107002;%%
\bibitem{gluodyn04}%
  \BibitemOpen
  \bibfield{author}{%
  \bibinfo {author} {\bibfnamefont{L.}~\bibnamefont{Li}}\ and\ \bibinfo
  {author} {\bibfnamefont{Y.}~\bibnamefont{Meurice}},\ }%
  \bibfield{journal}{%
  \bibinfo {journal} {Phys. Rev. D}\ }%
  \textbf{\bibinfo {volume} {71}},\ \bibinfo {pages} {016008} (\bibinfo {year}
  {2005}),\ \Eprint{http://arxiv.org/abs/hep-lat/0410029}{hep-lat/0410029}%
  \bibAnnoteFile{NoStop}{gluodyn04}%
%%CITATION = HEP-LAT 0410029;%%
\bibitem{Denbleyker:2008ss}%
  \BibitemOpen
  \bibfield{author}{%
  \bibinfo {author} {\bibfnamefont{A.}~\bibnamefont{Denbleyker}}, \bibinfo
  {author} {\bibfnamefont{D.}~\bibnamefont{Du}}, \bibinfo {author}
  {\bibfnamefont{Y.}~\bibnamefont{Liu}}, \bibinfo {author}
  {\bibfnamefont{Y.}~\bibnamefont{Meurice}},\ and\ \bibinfo {author}
  {\bibfnamefont{A.}~\bibnamefont{Velytsky}},\ }%
  \bibfield{journal}{%
  \Doi{10.1103/PhysRevD.78.054503}{\bibinfo {journal} {Phys. Rev. D}}\ }%
  \textbf{\bibinfo {volume} {78}},\ \bibinfo {pages} {054503} (\bibinfo {year}
  {2008}),\ \Eprint{http://arxiv.org/abs/0807.0185}{arXiv:0807.0185 [hep-lat]}%
  \bibAnnoteFile{NoStop}{Denbleyker:2008ss}%
%%CITATION = 0807.0185;%%
\bibitem{balian74err}%
  \BibitemOpen
  \bibfield{author}{%
  \bibinfo {author} {\bibfnamefont{R.}~\bibnamefont{Balian}}, \bibinfo {author}
  {\bibfnamefont{J.~M.}\ \bibnamefont{Drouffe}},\ and\ \bibinfo {author}
  {\bibfnamefont{C.}~\bibnamefont{Itzykson}},\ }%
  \bibfield{journal}{%
  \bibinfo {journal} {Phys. Rev. D}\ }%
  \textbf{\bibinfo {volume} {19}},\ \bibinfo {pages} {2514} (\bibinfo {year}
  {1979})%
  \bibAnnoteFile{NoStop}{balian74err}%
%%CITATION = PHRVA,D19,2514;%%
\bibitem{michael88}%
  \BibitemOpen
  \bibfield{author}{%
  \bibinfo {author} {\bibfnamefont{C.}~\bibnamefont{Michael}}\ and\ \bibinfo
  {author} {\bibfnamefont{M.}~\bibnamefont{Teper}},\ }%
  \bibfield{journal}{%
  \bibinfo {journal} {Nucl. Phys.}\ }%
  \textbf{\bibinfo {volume} {B314}},\ \bibinfo {pages} {347} (\bibinfo {year}
  {1989})%
  \bibAnnoteFile{NoStop}{michael88}%
%%CITATION = NUPHA,B314,347;%%
\bibitem{Horsley:1981gj}%
  \BibitemOpen
  \bibfield{author}{%
  \bibinfo {author} {\bibfnamefont{R.}~\bibnamefont{Horsley}}\ and\ \bibinfo
  {author} {\bibfnamefont{U.}~\bibnamefont{Wolff}},\ }%
  \bibfield{journal}{%
  \Doi{10.1016/0370-2693(81)90891-1}{\bibinfo {journal} {Phys. Lett.}}\ }%
  \textbf{\bibinfo {volume} {B105}},\ \bibinfo {pages} {290} (\bibinfo {year}
  {1981})%
  \bibAnnoteFile{NoStop}{Horsley:1981gj}%
%%CITATION = PHLTA,B105,290;%%
\bibitem{npp}%
  \BibitemOpen
  \bibfield{author}{%
  \bibinfo {author} {\bibfnamefont{Y.}~\bibnamefont{Meurice}},\ }%
  \bibfield{journal}{%
  \Doi{10.1103/PhysRevD.74.096005}{\bibinfo {journal} {Phys. Rev. D}}\ }%
  \textbf{\bibinfo {volume} {74}},\ \bibinfo {pages} {096005} (\bibinfo {year}
  {2006}),\
  \Eprint{http://arxiv.org/abs/hep-lat/0609005}{arXiv:hep-lat/0609005}%
  \bibAnnoteFile{NoStop}{npp}%
%%CITATION = HEP-LAT/0609005;%%
\bibitem{1991JSP....62..529B}%
  \BibitemOpen
  \bibfield{author}{%
  \bibinfo {author} {\bibfnamefont{C.}~\bibnamefont{{Borgs}}}, \bibinfo
  {author} {\bibfnamefont{R.}~\bibnamefont{{Koteck{\'y}}}},\ and\ \bibinfo
  {author} {\bibfnamefont{S.}~\bibnamefont{{Miracle-Sol{\'e}}}},\ }%
  \bibfield{journal}{%
  \Doi{10.1007/BF01017971}{\bibinfo {journal} {Journal of Statistical
  Physics}}\ }%
  \textbf{\bibinfo {volume} {62}},\ \bibinfo {pages} {529} (\bibinfo {month}
  {Feb.}\ \bibinfo {year} {1991})%
  \bibAnnoteFile{NoStop}{1991JSP....62..529B}%
\bibitem{1990JSP....61...79B}%
  \BibitemOpen
  \bibfield{author}{%
  \bibinfo {author} {\bibfnamefont{C.}~\bibnamefont{{Borgs}}}\ and\ \bibinfo
  {author} {\bibfnamefont{R.}~\bibnamefont{{Koteck{\'y}}}},\ }%
  \bibfield{journal}{%
  \Doi{10.1007/BF01013955}{\bibinfo {journal} {Journal of Statistical
  Physics}}\ }%
  \textbf{\bibinfo {volume} {61}},\ \bibinfo {pages} {79} (\bibinfo {month}
  {Oct.}\ \bibinfo {year} {1990})%
  \bibAnnoteFile{NoStop}{1990JSP....61...79B}%
\bibitem{Billoire:1992ke}%
  \BibitemOpen
  \bibfield{author}{%
  \bibinfo {author} {\bibfnamefont{A.}~\bibnamefont{Billoire}}, \bibinfo
  {author} {\bibfnamefont{T.}~\bibnamefont{Neuhaus}},\ and\ \bibinfo {author}
  {\bibfnamefont{B.}~\bibnamefont{Berg}},\ }%
  \bibfield{journal}{%
  \Doi{10.1016/0550-3213(93)90671-B}{\bibinfo {journal} {Nucl. Phys.}}\ }%
  \textbf{\bibinfo {volume} {B396}},\ \bibinfo {pages} {779} (\bibinfo {year}
  {1993}),\
  \Eprint{http://arxiv.org/abs/hep-lat/9211014}{arXiv:hep-lat/9211014}%
  \bibAnnoteFile{NoStop}{Billoire:1992ke}%
%%CITATION = HEP-LAT/9211014;%%
\bibitem{Boyd:1988}%
  \BibitemOpen
  \bibfield{author}{%
  \bibinfo {author} {\bibfnamefont{J.~P.}\ \bibnamefont{Boyd}},\ }%
  \bibfield{journal}{%
  \bibinfo {journal} {Journal of Scientific Computing}\ }%
  \textbf{\bibinfo {volume} {3}},\ \bibinfo {pages} {109} (\bibinfo {year}
  {1998})%
  \bibAnnoteFile{NoStop}{Boyd:1988}%
\bibitem{124403}%
  \BibitemOpen
  \bibfield{author}{%
  \bibinfo {author} {\bibfnamefont{P.}~\bibnamefont{Kravanja}}, \bibinfo
  {author} {\bibfnamefont{M.}~\bibnamefont{Van~Barel}}, \bibinfo {author}
  {\bibfnamefont{O.}~\bibnamefont{Ragos}}, \bibinfo {author}
  {\bibfnamefont{M.}~\bibnamefont{Vrahatis}},\ and\ \bibinfo {author}
  {\bibfnamefont{F.}~\bibnamefont{Zafiropoulos}},\ }%
  \bibfield{journal}{%
  \bibinfo {journal} {Computer Physics Communications}\ }%
  \textbf{\bibinfo {volume} {124}},\ \bibinfo {pages} {212} (\bibinfo {month}
  {Feb.}\ \bibinfo {year} {2000})%
  \bibAnnoteFile{NoStop}{124403}%
\bibitem{BBbook}%
  \BibitemOpen
  \bibfield{author}{%
  \bibinfo {author} {\bibfnamefont{B.}~\bibnamefont{Berg}},\ }%
  \emph{\bibinfo {title} {Markov Chain Monte Carlo Simulations and Their
  Statistical Analysis}}\ (\bibinfo {publisher} {World Scientific},\ \bibinfo
  {address} {Singapore},\ \bibinfo {year} {2004})%
  \bibAnnoteFile{NoStop}{BBbook}%
\bibitem{PhysRevE.74.011108}%
  \BibitemOpen
  \bibfield{author}{%
  \bibinfo {author} {\bibfnamefont{H.}~\bibnamefont{Behringer}}\ and\ \bibinfo
  {author} {\bibfnamefont{M.}~\bibnamefont{Pleimling}},\ }%
  \bibfield{journal}{%
  \Doi{10.1103/PhysRevE.74.011108}{\bibinfo {journal} {Phys. Rev. E}}\ }%
  \textbf{\bibinfo {volume} {74}},\ \bibinfo {pages} {011108} (\bibinfo {month}
  {Jul}\ \bibinfo {year} {2006})%
  \bibAnnoteFile{NoStop}{PhysRevE.74.011108}%
\bibitem{wilson74c}%
  \BibitemOpen
  \bibfield{author}{%
  \bibinfo {author} {\bibfnamefont{K.~G.}\ \bibnamefont{Wilson}},\ }%
  \bibfield{journal}{%
  \bibinfo {journal} {Phys. Rev. D}\ }%
  \textbf{\bibinfo {volume} {10}},\ \bibinfo {pages} {2445} (\bibinfo {year}
  {1974})%
  \bibAnnoteFile{NoStop}{wilson74c}%
%%CITATION = PHRVA,D10,2445;%%
\bibitem{Berg:2006hh}%
  \BibitemOpen
  \bibfield{author}{%
  \bibinfo {author} {\bibfnamefont{B.~A.}\ \bibnamefont{Berg}}\ and\ \bibinfo
  {author} {\bibfnamefont{A.}~\bibnamefont{Bazavov}},\ }%
  \bibfield{journal}{%
  \Doi{10.1103/PhysRevD.74.094502}{\bibinfo {journal} {Phys. Rev. D}}\ }%
  \textbf{\bibinfo {volume} {74}},\ \bibinfo {pages} {094502} (\bibinfo {year}
  {2006}),\
  \Eprint{http://arxiv.org/abs/hep-lat/0605019}{arXiv:hep-lat/0605019}%
  \bibAnnoteFile{NoStop}{Berg:2006hh}%
%%CITATION = HEP-LAT/0605019;%%
\end{thebibliography}
%\end{document}
%Merlin.mbs v4.21 2009-07-09.

%\newpage
%

\end{document}